\documentclass[a4paper,twocolumn,aps,prb,superscriptaddress,showpacs,floatfix]{revtex4}
\usepackage{amssymb,amsmath}

\usepackage{array}
\usepackage{graphicx}
\usepackage{ifpdf}
\usepackage{color} 
\usepackage{url}
\usepackage{hyperref}
\definecolor{darkblue}{rgb}{0.0,0.0,0.3}
\hypersetup{colorlinks,breaklinks,
            linkcolor=blue,urlcolor=blue,
            anchorcolor=blue,citecolor=blue}

\renewcommand{\vec}[1]{\mathbf{#1}}

\mathchardef\mhyphen="0002D

\renewcommand{\Re}{\text{Re}\,}

\renewcommand{\Im}{\text{Im}\,}
\newcommand{\plus}{ {\scriptscriptstyle +}}
\newcommand{\minus}{{\scriptscriptstyle -}}

\newcommand{\ket}{\rangle}
\newcommand{\bra}{\langle}
\newcommand{\be}{\begin{equation}}
\newcommand{\ee}{\end{equation}}
\newcommand{\bea}{\begin{eqnarray}}
\newcommand{\eea}{\end{eqnarray}}

\newcommand{\mC}{\mathcal{C}}
\newcommand{\mF}{\mathcal{F}}

\newcommand{\mU}{\mathcal{U}}

\newcommand{\hmU}{\hat{\mathcal{U}}}

\newcommand{\hg}{\hat{\gamma}}
\newcommand{\hgd}{\hat{\gamma}^{\dagger}}

\newcommand{\hH}{\hat{H}}

\newcommand{\hpsi}{\hat{\psi}}
\newcommand{\hpsid}{\hat{\psi}^{\dagger}}

\newcommand{\br}{\mathbf{r}}

\newcommand{\bx}{\mathbf{x}}
\newcommand{\by}{\mathbf{y}}

\newcommand{\pu}{\underline{p}}
\newcommand{\qu}{\underline{q}}

\newcommand{\s}{\sigma}

\newcommand{\w}{\omega}

\newcommand{\nn} {\nonumber}

\newcommand{\Si} { \Sigma}

\def\a{\alpha}
\def\b{\beta}

\def\s{\sigma}

\def\e{\epsilon}
\def\t{\tau}

\usepackage{lipsum}

\begin{document}
%-----------------------------------------------------------------------------------------

\title{Diagrammatic expansion for positive spectral functions beyond GW: Application to vertex corrections in the electron gas}

\author{G. Stefanucci}
\affiliation{Dipartimento di Fisica, Universit{\`a} di Roma Tor Vergata, 
Via della Ricerca Scientifica 1, 00133 Rome, Italy}
\affiliation{INFN, Laboratori Nazionali di Frascati, Via E. Fermi 40,
00044 Frascati, Italy}
\affiliation{European Theoretical Spectroscopy Facility (ETSF)}

\author{Y. Pavlyukh}
\affiliation{Institut f\"{u}r Physik, Martin-Luther-Universit\"{a}t
  Halle-Wittenberg, 06120 Halle, Germany}

\author{A.-M. Uimonen}
\affiliation{Department of Physics, Nanoscience Center, University of Jyv{\"a}skyl{\"a}, 
FI-40014 Jyv{\"a}skyl{\"a}, Finland}

\author{R. van Leeuwen}
\affiliation{Department of Physics, Nanoscience Center, University of Jyv{\"a}skyl{\"a}, 
FI-40014 Jyv{\"a}skyl{\"a}, Finland}
\affiliation{European Theoretical Spectroscopy Facility (ETSF)}
\date{\today}
%*****************************************************************************************
\begin{abstract}

We present a diagrammatic approach to construct self-energy approximations
within many-body perturbation theory with positive spectral 
properties. The method cures
the problem of negative spectral functions which arises from a straightforward inclusion
of vertex diagrams beyond the GW approximation. Our approach 
consists of a two-steps 
procedure: we first express the approximate many-body self-energy as a product of
half-diagrams and then identify the minimal number of half-diagrams 
to add in order to form a perfect square. The resulting self-energy 
is an unconventional sum of self-energy diagrams in which 
the internal lines of half a diagram are time-ordered  Green's functions
whereas those of the other half are anti-time-ordered 
Green's functions, and the lines joining the two halves are either 
lesser or greater Green's functions.
% expression for the spectral function of the self-energy in terms of Keldysh
% diagrams. Using the Langreth rules we derive a Lehmann representation for this spectral
% function and give a diagrammatic representation of this expression as a product of
% half-diagrams.  The positivity of a given self-energy approximation can be inferred by
% collecting the half-diagrams into a full square. 
% If a given self-energy approximation is
% already a complete square of half-diagrams, the self-energy will yield positive
% spectra. For other self-energy approximations we will need to add specific missing terms
% in order to be able to complete the square.  In these situations the given self-energy
% approximation will not give positive spectra without the addition of these extra
% terms. 
The theory is developed using noninteracting Green's 
functions and subsequently extended to self-consistent Green's 
functions. Issues related to the conserving properties of 
diagrammatic approximations with positive 
spectral functions are also addressed. As a major application of the 
formalism we derive the minimal set of additional diagrams to
make positive the spectral function of the GW approximation with 
lowest-order vertex corrections and screened interactions.  The method is then applied to vertex corrections in the three-dimensional
homogeneous electron gas by using a combination of analytical frequency integrations and
numerical Monte-Carlo momentum integrations to evaluate the diagrams.

\end{abstract}
\pacs{71.10.-w,31.15.A-,73.22.Dj}
%71.10.-w 	Theories and models of many-electron systems
%31.15.A- 	Ab initio calculations
%73.22.Dj 	Single particle states
%\keywords{}
\maketitle
%*****************************************************************************************
\section{Introduction}
%*****************************************************************************************

Many-body perturbation theory (MBPT) has provided a systematic way to study
electron-electron (electron-phonon) interactions in various systems ranging from molecules
to solids.~\cite{kadanoff_quantum_1962,stefanucci_nonequilibrium_2013} Within MBPT the interaction effects are included
via a self-energy term which is treated perturbatively.  One of the widely used
self-energy approximations is the GW approximation~\cite{hedin_new_1965} which consists of
replacing the bare interaction with the screened interaction in the 
first-order exchange diagram.
Diagrammatically the GW approximation can be viewed as an infinite
summation of polarization diagrams. It is well known that for 
solids the GW approximation (usually not implemented self-consistently)
tends to give band gap values close to the experimental 
values, thus improving
over the density functional calculations (which instead underestimate the values for the band
gaps).\cite{aulbur} In spite of some improvements over complementary theories, the self-consistent GW
approximation is known to have a number of deficiencies like the washing out of plasmon features and
broadened bandwidths in the electron-gas-like 
metals.~\cite{holm_fully_1998} For many
decades the common argument has then been that the inclusion of vertex corrections would
act as a balancing force for the 
self-consistency,~\cite{mahan_vertex_1994,hong_mahan_1994,verdozzi,degroot}
thus, e.g., hampering the washing out of plasmon satellites.  Several people have  worked
on this issue on various levels,~\cite{minnhagen_vertex_1974,
  del_sole_gw_1994,shirley_self-consistent_1996,ness_gw_2011,takada_inclusion_2001,takada_dynamical_2002,romaniello}
but the most interesting result from our point of view is that the 
straightforward inclusion of
vertex corrections beyond the GW level yields negative spectra in 
some frequency regions, as first
noticed by Minnhagen for the electron 
gas.~\cite{minnhagen_vertex_1974}  This 
deficiency not only prohibits the usual probability interpretation of the
spectral function but also generates Green's
functions with the wrong analytic properties. In particular the latter feature prevents an iterative self-consistent solution
of the Dyson equation since the analytic properties deteriorate with every self-consistency
cycle.  This unpleasant situation is not limited to the electron gas as it has also been observed in a
study of vertex corrections in finite 
systems.~\cite{schindlmayr_hubbard98,arno} As we will explain in depth in this work the
problem lies in the structure of the vertex correction and therefore 
the solution must be sought in the way we use MBPT.  There has been 
very little work on
how to generate positive spectral functions from MBPT.  The only
work on the issue of positivity that we are aware of has been done by 
Almbladh but in the context of photoemission. Almbladh showed that
the positivity of the photocurrent was
guaranteed by expressing the photo-emission
triangle diagrams as a square of half-diagrams.~\cite{almbladh_photoemission_2006} 
This, however, was done only for a certain selection of low-order
diagrams and it was not indicated how the idea
could be applied to the spectral function.

In this paper we put forward a diagrammatic approach to generate self-energy approximations
beyond GW yielding positive spectral functions.  We start from an 
expression of the self-energy derived by 
Danielewicz~\cite{danielewicz_quantum_1984} and use the Keldysh
formalism~\cite{keldysh_diagram_1965} to extract the lesser/greater 
components (these components are needed to construct the spectral 
function). Every lesser/greater diagram is partitioned into different 
contributions, each with internal times integrated over either the 
forward or the backward branch of 
the Keldysh contour. The full lesser/greater diagram corresponds to 
the sum of all possible partitions. We then factorize each partition 
into half-diagrams by using the Lehmann
representation of the Green's function,~\cite{kallen_definition_1952,lehmann_uber_1954} 
where the one half of the
partition consists of time-ordered quantities and the other half 
consists of
anti-time-ordered quantities. The partitioning can be seen as cutting the diagram in half
along the lesser/greater Green's function lines. 
A similar cutting procedure is used in high-energy physics to calculate the imaginary
part of diagrams contributing to the scattering
amplitudes.~\cite{cutkosky_singularities_1960,veltman_unitarity_1963,
kobes_discontinuities_1985,kobes_discontinuities_1986,kobes_cutkosky_1986,jeon_computing_1993}
Our cutting rules agree with those of high-energy physics but our 
derivation is based on the Keldysh formalism, which allows us to advance 
the theory further. In fact, the positivity of 
the spectral function entails that the sum of the products of the
half-diagrams is the sum of perfect squares.
 In some situations the MBPT
approximation is already a sum of perfect squares, 
like for the GW approximation.
For other self-energy approximations we instead need to add missing 
half-diagrams to complete the square, like for the $GWGGW$ 
self-energy, i.e., the lowest-order vertex correction. We acknowledge 
here that Danielewicz~\cite{danielewicz_operator_1990} also studied a cutting
procedure for the lesser/greater self-energy diagrams and derived a 
manifestly positive exact formula  for the spectral function
%for a general time-ordered $n$-point function 
%with the aim of deriving general scattering rate formulae 
in terms of retarded/advanced $n$-point functions, 
but the issue of how to cure negative
spectral functions of 
approximate self-energies  was not discussed in his work.  
The focus of our work is to study approximate MBPT
self-energies and give simple drawing rules to decide whether or not 
the approximation generate a
positive spectral function.  If not we provide extra, but still simple, drawing rules 
to extend to a minimal set of diagrams the MBPT approximation and 
turn positive the spectral function.

This paper is organized as follows. In section~\ref{sec:theory} we derive the Lehmann form
of the lesser/greater self-energy and relate it to a diagrammatic representation in terms of
half-diagrams. Then we describe how to construct a self-energy approximation with a
positive spectral function from a given MBPT approximation by a minimal selection of additional
self-energy diagrams.  The theory is developed using noninteracting 
Green's functions and subsequently extended to self-consistent 
Green's functions. In section~\ref{examples} we illustrate the 
formalism  with text-book examples. In section~\ref{beyondGW} we derive the
simplest self-energy with {\em vertex corrections}
and {\em screened interaction} yielding a positive spectral function.  
We then apply the theory to the three-dimensional homogeneous electron gas 
in section~\ref{sec:numerics}. We evaluate the self-energy diagrams   
by using a combination of analytical frequency integrations and numerical Monte-Carlo
momentum integrations, and show how the minimal selection of additional 
diagrams cures the problem of negative spectra.  We finally present 
our conclusions and outlooks in section~\ref{c&osec}.

%*****************************************************************************************
\section{Theoretical Framework\label{sec:theory}}
%*****************************************************************************************

\begin{figure}[b!]
	\centering
       \includegraphics[width=0.9\linewidth]{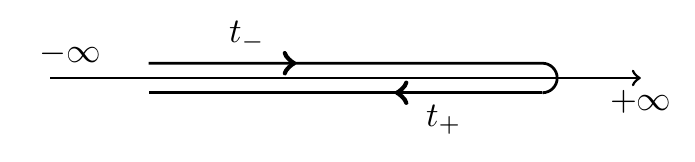}
       \caption[]{The closed time-loop contour $\mC$. The forward branch is
       denoted with a ``$-$'' label while the backward branch is denoted 
       by a ``$+$'' label.}
      \label{fig:contour}
\end{figure}

Within the realm of Green's function theory the most common techniques to study
equilibrium problems are either the zero-temperature formalism or the Matsubara
formalism. These are two special cases of the more general Keldysh formalism which is
usually applied in the context of non-equilibrium physics beyond linear response. In this
work we show that the Keldysh formalism is also the natural tool to develop a diagrammatic
theory for {\em positive-definite} spectral functions of systems in {\em equilibrium}.  We
consider a system of interacting fermions with Hamiltonian
%===========================================================
%
%===========================================================
\bea 
\hat{H}&=&\int d\bx
\,\hpsid(\bx)h(\bx)\hpsi(\bx) \nn\\ &+&\frac12\int d\bx d\bx'
\hpsid(\bx)\hpsid(\bx')v(\bx,\bx')\hpsi(\bx')\hpsi(\bx),
\label{ham}
\eea 
%===========================================================
%
%===========================================================
where the field operator $\hpsi$ ($\hpsi^{\dag}$) with argument 
$\bx=\br\s$ annihilates
(creates) a fermion in position $\br$ with spin $\s$. In the Keldysh formalism
the field operators evolve on the time-loop contour $\mC$ shown in Fig.~\ref{fig:contour}.
Operators on the {\em minus}-branch are ordered chronologically while operators on the
{\em plus}-branch are ordered anti-chronologically.
%---------------------------
%---------------------------
Letting $z_{1}$ and $z_{2}$ be two contour-times, the Green's function $G(\bx_1z_1,\bx_2
z_2)$ can be divided into different components $G^{\a\b}(\bx_1t_1,\bx_2 t_2)$ depending on
the branch $\a,\b=+/-$ to which $z_{1}$ and $z_{2}$ belong. For $\a=\b=-$ we have the
\emph{time-ordered} Green's function 
%===========================================================
%
%===========================================================
\be 
G^{\minus\minus} (\bx_1t_1,\bx_2 t_2) = -i \bra T
\left[ \hpsi_H (\bx_1 t_1) \hpsid_H (\bx_2 t_2)\right] \ket. 
\label{eq:gmm}
\ee 
%===========================================================
%
%===========================================================
In this expression the
average $\bra\ldots \ket$ is done with some density matrix $\hat{\rho}$ and $T$ is the
time-ordering operator.  The subscript ``$H$'' attached to a general operator $\hat{O}$
signifies that that operator is in the Heisenberg picture 
%===========================================================
%
%===========================================================
\be
\hat{O}_{H}(t)=\hat{\mU}(t_0,t)\hat{O}\,\hat{\mU}(t,t_{0}),
\label{hp}
\ee 
%===========================================================
%
%===========================================================
where $\hat{\mU}(t_1,t_2)$ is the time-evolution operator and $t_{0}$ is an arbitrary
initial time. Reversing the time arrow the $G^{\minus\minus}$ is converted into the
\emph{anti-time-ordered} Green's function 
%===========================================================
%
%===========================================================
\be 
G^{\plus\plus} (\bx_1 t_1,\bx_2 t_2) = -i
\bra \bar{T} \left[ \hpsi_H (\bx_1 t_1) \hpsid_H (\bx_2 t_2)\right] \ket, 
\ee 
%===========================================================
%
%===========================================================
where
$\bar{T}$ orders the operators anti-chronologically. Finally, choosing $z_{1}$ and $z_{2}$
on different branches we have
%===========================================================
%
%===========================================================
\begin{subequations}
\label{g><}
\bea G^{\minus\plus} (\bx_1 t_1,\bx_2 t_2) &=&  i \bra
\hpsid_H (\bx_2t_2) \hpsi_H (\bx_1 t_1) \ket,
\label{g<}\\
G^{\plus\minus} (\bx_1 t_1,\bx_2t_2) &=&-i \bra \hpsi_H (\bx_1 t_1) \hpsid_H (\bx_2 t_2) \ket.
\label{g>}
\eea 
\end{subequations}
%===========================================================
%
%===========================================================
These two last components are equivalently written as $G^{\minus\plus}=G^{<}$
(\emph{lesser} Green's function) and $G^{\plus\minus}=G^{>}$ (\emph{greater} Green's
function), and describe the propagation of an added hole ($G^{<}$) or particle ($G^{>}$)
in the medium.

%**************************************************************
\subsection{Positive Semidefiniteness of the Exact Self-Energy}
\label{possemidef}
%**************************************************************

In equilibrium the Green's function $G^{\a\b}$ depends on the time-difference
only. Omitting the dependence on the position and spin coordinates $\bx_{1}$ and $\bx_{2}$
the spectral function is defined according to 
\be
A(\w)=i[G^{>}(\w)-G^{<}(\w)],
\ee
where here
and in the following $G^{\a\b}(\w)$ denotes the Fourier transform of the Green's function
with respect to the time-difference. From the Lehmann representation it is easy to show
that $iG^{>}(\w)$ and $-iG^{<}(\w)$, as matrices in the $\bx$-space, are positive
semidefinite (PSD). From the Dyson equation on the Keldysh contour 
one can also show that\cite{stefanucci_nonequilibrium_2013}
\be
G^{\lessgtr}(\w)=G^{\rm R}(\w)\Sigma_{c}^{\lessgtr}(\w)G^{\rm A}(\w),
\ee
where 
\be G^{\rm
  R/A}(\w)=i\int\frac{d\w'}{2\pi}\frac{G^{>}(\w')-G^{<}(\w')}{\w-\w'\pm i\eta} 
  \label{GRA}
 \ee 
  are the
retarded/advanced Green's function and $\Sigma_{c}$ is the correlation self-energy. Since
$G^{\rm A}(\w)=[G^{\rm R}(\w)]^{\dag}$ the PSD of $\mp i G^{\lessgtr}$ implies that $\mp i
\Sigma_{c}^{\lessgtr}$ is PSD and vice versa. Even though one can prove the PSD property of
the self-energy using the corresponding property of the Green's function
and the Dyson equation, a direct proof starting from a Lehmann representation is not possible
since $\Sigma_{c}$ is {\em not} the average of a correlator.\\
%\begin{equation}
%\Sigma_c(\vec x_1,\vec x_2;\omega)=\sum_\nu\frac{m_\nu(\vec x_2)m^*_\nu(\vec x_1)}
%{\omega-\epsilon_{\nu}}
%\end{equation}
%is not possible because $m_\nu(\vec x_1)$ has no clear diagrammatic meaning and can only
%be defined as a solution of some linear equation.~\cite{winter_study_1972} In other words
In this section we use the Keldysh formalism to provide an alternative proof of the PSD
property of $\Sigma_{c}$. For the proof we derive a Lehmann-like representation of 
$\Sigma_{c}$ and highlight the 
connection with the diagrammatic expansion. This connection will be extremely useful to generate
approximate PSD self-energies from diagrammatic theory.
The starting point is the following expression\cite{danielewicz_quantum_1984,stefanucci_nonequilibrium_2013} for
$\Si_{c}^{<}$ and $\Si^{>}_{c}$ 
%===========================================================
%
%===========================================================
\begin{subequations}
\label{s<>}
\bea \Si_c^{<}(\bx_1 t_1,\bx_2t_2) &=& i\bra \hgd_H(\bx_2
t_2) \hg_H (\bx_1 t_1) \ket_{\rm irr}\;,
\label{s<}\\
\Si_c^{>}(\bx_1 t_1,\bx_2 t_2) &=&-i \bra \hg_H (\bx_1 t_1) 
\hgd_H(\bx_2t_2) \ket_{\rm irr}\;,
\label{s>}
\eea 
\end{subequations}
%===========================================================
%
%===========================================================
where the operator $\hg$ is defined according to \be \hg(\bx_1) \equiv \int d\bx_2
v(\bx_1,\bx_2) \hpsid(\bx_{2})\hpsi(\bx_{2})\hpsi(\bx_1), \ee and the subscript ``irr''
signifies that only one-particle irreducible diagrams should be retained. That is, the
expansion of the self-energy~\eqref{s<>} contains all the diagrams of the
two-particle-one-hole correlation function~\cite{pavlyukh_initial_2013} in
which the entrance and exit channels cannot be separated by cutting one Green's function
line.~\cite{ethofer_six-point_1969,winter_study_1972} Unlike the definitions (\ref{g><})
of the Green's functions $G^{\lessgtr}$, the equations \eqref{s<>} are not averages of a
correlator due to the exclusion of reducible diagrams. Nevertheless a Lehmann-like
representation for the self-energy can be derived using diagrammatic methods.

Let us study, e.g., $\Si_c^{<}$ as the same reasoning applies to $\Si^{>}_{c}$. For
simplicity we restrict the discussion to systems at zero temperature and assume a
nondegenerate ground state $\Psi_{0}$. Using Eq.~(\ref{hp}) and introducing the
short-hand notation $1=\bx_1 t_1$, $2=\bx_2 t_2$, etc. we can rewrite Eq.  (\ref{s>}) as
%===========================================================
%
%===========================================================
\bea
\Si_c^{<}(1,2)&=&i \bra \Psi_0 | \hmU(t_0,t_2)
\hgd(\bx_{2})\hmU(t_2,t_0)\nn\\ &&\hmU(t_0,t_1) \hg(\bx_{1}) \hmU(t_1,t_0)| \Psi_0
\ket_{\rm irr}\;.
\label{s<2}
\eea 
%===========================================================
%
%===========================================================
Next we assume that $\Psi_{0}$ can be obtained by evolving backward the noninteracting
ground state $\Phi_{0}$ from a distant future time $\t$ (with 
$\tau\to \infty$) to the arbitrary initial time $t_{0}$ using an
interaction which is switched-on adiabatically, i.e., $|\Psi_0
\ket = \hat{\mU} (t_0,\tau) | \Phi_0\ket$ where the evolution 
operator is calculated with the time-dependent interaction
$e^{\eta |t-t_0|} v$ ($\eta$ being an infinitesimal energy). This is the standard assumption of the 
zero-temperature Green's function formalism.  
%We also notice that sometimes it is convenient
%to include the mean field Coulomb interaction (the Hartree term) into the noninteracting
%Hamiltonian. In this case the self-energy is free from tadpole diagrams. 
Then
Eq.~\eqref{s<2} becomes (the limit $\t\to\infty$ is implied)
%===========================================================
%
%===========================================================
\bea
\Si_c^{<}(1,2)=i\bigg [ \sum_i \bra \Phi_0 |
  \hat{\mU}(\tau,t_2) \hgd(\bx_2) \hat{\mU}(t_2,\tau) |\chi_i \ket \nn\\ \times\bra
  \chi_i| \hat{\mU}(\tau,t_1) \hg(\bx_1) \hat{\mU}(t_1,\tau)| \Phi_0 
  \ket \bigg]_{\rm irr},
\label{s<3}
\eea
%===========================================================
%
%===========================================================
where we inserted a completeness relation $\sum_{i}|\chi_i \ket \bra \chi_i|=1$ and
used the group property $\hmU(t_1,t_0)\hmU(t_0,\tau)=\hmU(t_1,\tau)$ and
$\hmU^\dagger(t_0,\tau)\hmU(t_0,t_2)=\hmU(\tau,t_2)$.  Let $\hat{c}_k$,
$\hat{c}_k^{\dagger}$ denote the annihilation and creation operators of a fermion in the
$k$-th eigenstate of the noninteracting problem.  In Eq.~(\ref{s<3}) only states $|\chi_i
\ket$ of the form 
\be
\hat{c}_{q_N}^{\dagger}
\ldots\hat{c}_{q_1}^{\dagger}\hat{c}_{p_{N+1}}\ldots\hat{c}_{p_1} | \Phi_0 \ket \equiv
|\chi_{\pu\qu}^{(N)}\ket
\ee
contribute since the operator $\hg$ ($\hgd$) annihilates
(creates) a fermion. The indices $\pu=(p_{1},\ldots,p_{N+1})$ and
$\qu=(q_{1},\ldots,q_{N})$ in $\chi_{\pu\qu}^{(N)}$ specify the quantum numbers of the
$\hat{c}$ and $\hat{c}^{\dag}$ operators respectively. We conclude that Eq.~(\ref{s<3})
does not change under the replacement 
%===========================================================
%
%===========================================================
\begin{equation}
 \sum_{i} | \chi_i\ket \bra \chi_i | \rightarrow
\sum_{N=0}^{\infty} \frac{1}{(N+1)!N!}  \sum_{\pu\qu}| \chi_{\pu\qu}^{(N)} \ket \bra
\chi_{\pu\qu}^{(N)}|.
\end{equation}
%===========================================================
%
%===========================================================
Here the inner sum denotes integrations or summations over sets of $\pu$ and $\qu$ quantum
numbers with the restriction that $\pu$ integration runs over the occupied and $\qu$
integration runs over the unoccupied states, respectively. The prefactor stems from the
inner product of the intermediate states, i.e.,
%===========================================================
%
%===========================================================
\begin{equation*}
\bra\chi_{\pu\qu}^{(N)}|\chi_{\pu'\qu'}^{(N')}
\ket=\delta_{N,N'}\sum_{P\in \pi_{N+1}}\sum_{Q\in\pi_{N}}(-)^{P+Q}\delta_{P(\pu),\pu'}\delta_{Q(\qu),\qu'},
\end{equation*}
%===========================================================
%
%===========================================================
where $P$ and $Q$ run over all possible permutations of $N+1$ and $N$ indices with parities
$(-)^{P}$ and $(-)^{Q}$, respectively.  We further denoted the permutation group of $N$ elements
by $\pi_{N}$. Defining the amplitudes
%===========================================================
%
%===========================================================
\bea S_{N,\pu\qu}^*(1)
&\equiv& \bra \chi_{\pu\qu}^{(N)} | \hat{\mU}(\tau,t_1) \hat{\gamma}(\bx_1)
\hat{\mU}(t_1,\tau)| \Phi_0 \ket,
\nn \\ 
S_{N,\pu\qu}(2) &\equiv& \bra \Phi_0 |
\hat{\mU}(\tau,t_2) \hg^{\dagger}(\bx_2) \hat{\mU}(t_2,\tau) 
|\chi_{\pu\qu}^{(N)} \ket,
\nn
\eea 
%===========================================================
%
%===========================================================
the lesser self-energy takes the following compact form 
%===========================================================
%
%===========================================================
\bea
\label{sigma_lesser}
\Sigma_c^{<}(1,2) \!=\!
i\!\left[\sum_{N=0}^{\infty} \frac{1}{(N+1)!N!} 
 \sum_{\pu\qu}S_{N,\pu\qu}(2) S_{N,\pu\qu}^*(1) 
 \!\right]_{\rm irr}\!\!.\;\;
\eea
%===========================================================
%
%===========================================================

\begin{figure}[t]
\centering
       \includegraphics[width=\columnwidth]{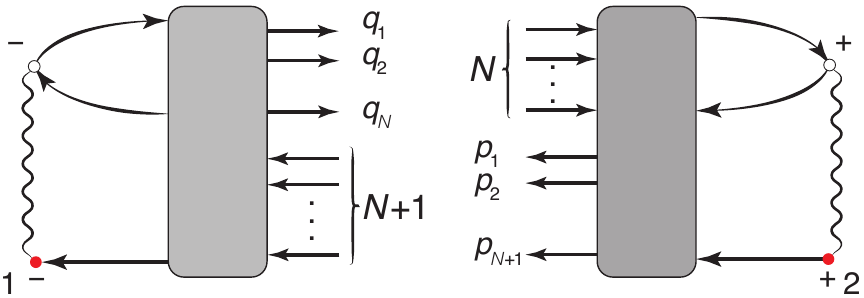}
\caption{(Color online) Diagrammatic structure of the functions $S(1)$ and $S^*(2)$ for 
the lesser self-energy. The external vertex points $1$ and $2$ have 
times on the $-$ and $+$ branch respectively. Green's
functions are denoted by the lines with arrows, while wavy lines correspond to the
bare interparticle interaction.
 \label{fig:ss}}
\end{figure}

To proceed further we need to analyze the amplitudes $S$ and their complex conjugate
$S^{\ast}$. Under the adiabatic assumption the evolution of the 
noninteracting ground state $\Phi_{0}$ from $-\tau$ to
$\tau$ yields $\Phi_{0}$ up to a phase factor, i.e., 
\be
\hmU (\tau,-\tau) | \Phi_0
\ket=e^{i\alpha}| \Phi_0 \ket
\ee
with $e^{i\alpha}=\bra\Phi_0 |\hmU (\tau,-\tau) | \Phi_0
\ket$.  Therefore we can write
%===========================================================
%
%===========================================================
\begin{widetext}
\bea
\label{s1}
S_{N,\pu\qu}^*(1)&=&
\bra\Phi_{0}|\hat{c}_{p_1}^{\dagger}\ldots
\hat{c}_{p_{N+1}}^{\dagger}\hat{c}_{q_1}\ldots \hat{c}_{q_N}
 \hat{\mU}(\tau,t_1) \hat{\gamma}(\bx_1)
 \hat{\mU}(t_1,-\tau)| \Phi_0 \ket \times e^{-i\a}
\nn\\
&=&\frac{\bra \Phi_0 | T \big\{  e^{-i\int_{-\tau}^{\tau} dt 
\hH(t)}\hat{c}_{p_1}^{\dagger}(\tau^+)\ldots
\hat{c}_{p_{N+1}}^{\dagger}(\tau^+)\hat{c}_{q_1}(\tau)\ldots\hat{c}_{q_N}(\tau) \hat{\gamma}(\bx_1t_1) \big\} |\Phi_0 \ket}
{\bra \Phi_0 |T \big\{  e^{-i\int_{-\tau}^{\tau} d t \hH(t)}  \big 
\} | \Phi_0 \ket},
\eea
%===========================================================
%
%===========================================================
where the time-argument in the operators specifies their position along the interval
$(-\tau,\tau)$. The time $\tau^+$ is infinitesimally greater than $\tau$, which assures the correct
ordering of the operators. The amplitude $S^*$ can now be expanded in powers of the
inter-particle interaction $v$ by means of Wick's theorem, and the generic term of the
expansion is a connected diagram of noninteracting {\em time ordered} Green's functions
$g^{\minus\minus}$ with external vertices $1=\bx_{1}t_{1}$ and $\pu$, $\qu$ at time
$\tau$, see left diagram in Fig.~\ref{fig:ss}. Following the same steps it is easy to show
that
%===========================================================
%
%===========================================================
\be
S_{N,\pu\qu}(2)=\frac{\bra\Phi_0 | \bar{T}  \big\{  
e^{i\int_{-\tau}^{\tau} d\tau \hH(t)}
\hat{\gamma}^{\dagger}(\bx_2t_2) \hat{c}_{q_N}^{\dagger}(\tau)\ldots\hat{c}_{q_1}^{\dagger}(\tau)
\hat{c}_{p_{N+1}}(\tau^+)\ldots\hat{c}_{p_1}(\tau^+)  \big \}| \Phi_0 \ket}{
 \bra \Phi_0 | \bar{T} \big\{  e^{i\int_{-\tau}^{\tau} dt \hH(t)}  \big \} |\Phi_0 \ket} ,
\label{Spq_def}
\ee
%===========================================================
%
%===========================================================
\end{widetext}
which can also be expanded using Wick's theorem, and the generic term of the expansion is
a connected diagram of noninteracting {\em anti-time ordered} Green's functions $g^{\plus\plus}$
with external vertices $2=\bx_{2}t_{2}$ and $\pu$, $\qu$ at time $\tau$, see right diagram
in Fig.~\ref{fig:ss}. Let us investigate the result of multiplying a diagram of
$S_{N,\pu\qu}^*(1)$ by a diagram of $S_{N,\pu\qu}(2)$ and then sum over $\pu$ and $\qu$.
In a diagram of $S^{\ast}_{N,\pu\qu}$ the outgoing Green's functions with $q$-labels and
the ingoing Green's functions with $p$-labels are calculated at the latest possible time
$\tau$. Therefore
%===========================================================
%
%===========================================================
\bea
g_{q\bx}^{\minus\minus}(\tau,t_{x}) &=&g_{q \bx}^{>}(\tau,t_{x}),
\nn\\
g_{\bx p}^{\minus\minus}(t_{x},\tau) &=& g_{\bx p}^{<}(t_{x},\tau),
\nn
\eea
%===========================================================
%
%===========================================================
where $(\bx,t_{x})$ is an internal space-spin-time vertex and we introduced the short-hand
notation $g_{ij}$ for the matrix elements of $g$ between two spin-orbital states $i$ and
$j$, hence $g^{\a\b}({\bx_{1}t_{1},\bx_{2}}t_{2})=g^{\a\b}_{\bx_{1}\bx_{2}}(t_{1},t_{2})$.
Similarly in a diagram of $S_{N,\pu\qu}$ the ingoing Green's functions with $q$-labels and
the outgoing Green's functions with $p$-labels are calculated at the latest possible time
$\tau$ and therefore
%===========================================================
%
%===========================================================
\bea
g_{\by q}^{\plus\plus}(t_{y},\tau) &=& g_{\by q}^{>}(t_{y},\tau),
\nn\\
g_{p\by}^{\plus\plus}(\tau,t_{y}) &=& g_{p\by}^{<}(\tau,t_{y}). \nn
\eea
%===========================================================
%
%===========================================================
%----------------------------------------
%---------------------------------------
\noindent
Thus the multiplication of a $S$ diagram by a $S^{\ast}$ diagram and the subsequent sum
over $\pu$ and $\qu$ involves the sum over $q$ of $g_{q \bx}^{>}g_{\by q}^{>}$ and the sum
over $p$ of $g_{\bx p}^{<}g_{p\by}^{<}$. The noninteracting lesser/greater Green's
functions can be expanded in the basis of the noninteracting one-particle eigenstates
according to\cite{stefanucci_nonequilibrium_2013}
%===========================================================
%
%===========================================================
\bea
g^{<}_{ij}(t,t')&=& i \sum_p f(\e_p)e^{-i\e_p(t-t')} \bra i | p \ket 
\bra p |j \ket, \label{gless_exp}\\
g^{>}_{ij}(t,t')&=& - i\sum_q \bar{f}(\e_q)e^{-i\e_q(t-t')} \bra i | 
q \ket \bra q |j \ket, \label{ggrt_exp}
\eea
%===========================================================
%
%===========================================================
where $\e_{p}$ is the energy of the one-particle eigenstate $|p\ket$, $f$ is the zero-temperature Fermi function
and $\bar{f}= 1-f$.  Taking into account that $f^{2}(\e_{p})=f(\e_{p})$ and
$\bar{f}^{2}(\e_{q})=\bar{f}(\e_{q})$ one can easily verify that
%===========================================================
%
%===========================================================
\bea \sum_q g_{\by
  q}^{>}(t_y,\t)g_{q\bx }^{>}(\t,t_{x})&=&-ig^{>}_{\by\bx}(t_y,t_x) ,\label{idemp_1} \\ 
  \sum_p g_{\bx
  p}^{<}(t_{x},\tau)g_{p\by}^{<}(\tau,t_{y})&=& ig_{\bx\by}^{<}(t_{x},t_{y})\label{idemp_2}.  
\eea
%===========================================================
%
%===========================================================
When taking the product of the left and the right half-diagram in 
Fig.~\ref{fig:ss} we can
use relation (\ref{idemp_1}) to replace each of the $N$ products of $g^>$ functions by a
single $g^>$ connecting two internal times in each half-diagram. Similarly we can use
relation (\ref{idemp_2}) to replace each of the $N+1$ products of a $g^<$ by a single
$g^<$. The result of this gluing procedure is a diagram with external
vertices $1$ and $2$.  The structure of the diagram is such that all internal vertices to
the left of the glued lines have time labels on the minus branch and 
all internal vertices to the right of the glued lines have time labels on
the plus branch. The gluing procedure can be reversed by cutting all Green's function
lines between a vertex labeled $-$ and a vertex labeled $+$. We will refer to this
procedure as the {\em cutting rule} for a diagram.  

\begin{figure}[t]
\centering
 \includegraphics[]{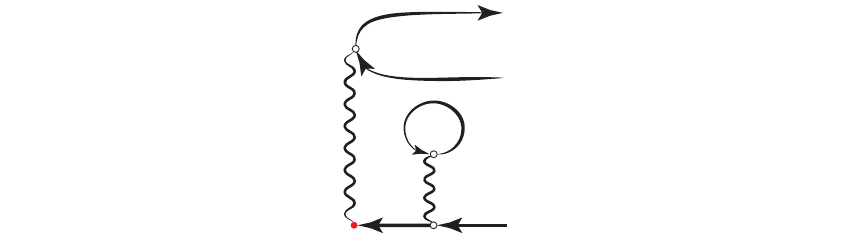}
\caption{(Color online) Minimal example of
  irreducible $S$ diagram with self-energy insertions. When glued with some irreducible $S^{\ast}$ diagram
  it yields an irreducible self-energy diagram. Notice that this diagram is reducible in the
  ``correlation function'' sense because cutting a single Green's 
  function line produces disconnected
  pieces.
 \label{fig:irrS}}
\end{figure}

At this point we observe
that the self-energy in Eq.~(\ref{sigma_lesser}) is not the sum of 
all possible $S$-$S^{\ast}$ diagrams due to the 
subscript ``irr''. This means
that from the diagrams obtained by the gluing procedure we still have the
remove the reducible diagrams, i.e., the diagrams which fall apart in two disjoint pieces
by cutting a single Green's function line. 
An obvious case of a reducible diagram is when there is only a single line to
glue in Fig.~\ref{fig:ss}.  This happens when $N=0$ and we only have the single label
$p_1$. This case is easily taken care of by letting the sum in Eq.~\eqref{sigma_lesser}
start at $N=1$ instead. For $N>1$ the gluing procedure leads to 
reducible diagrams whenever the $S$-diagram can be disjoint into a 
piece which contains {\em only} vertex 1 and a piece which contains the 
$\pu\qu$ vertices by cutting a single Green's function line. 
We call these $S$-diagrams reducible and define $\tilde{S}$ as
the sum of irreducible $S$-diagrams. Note that the $\tilde{S}$-diagrams 
{\em can} be disjoint into two pieces by cutting a single Green's 
function line, but then one of the pieces would contain 1 {\em and} some 
of the $\pu\qu$ vertices. For instance half-diagrams with a self-energy insertion 
on the lines we glue together belong to  $\tilde{S}$. An example is 
shown in Fig.~\ref{fig:irrS} where a tadpole in inserted in one of 
the $p$-lines. This half-diagram can be disjoint by a single cut but 
then the vertex 1 would not be isolated from all the  $\pu\qu$. 
Consequently this half-diagram belongs to $\tilde{S}$ and produces 
irreducible diagrams for the self-energy.
%Thus gluing two $g$'s yields a $g$. In view of this fact it is clear now that internal
%lines in a self-energy diagram (labeled as $\pu$ and $\qu$) are just external legs of the
%$S_{N,\pu\qu}^*(1)$ and $S_{N,\pu\qu}(2)$ correlation functions. These correlators depend
%on one time argument and $2N$ spatial coordinates and fulfill standard sign rules for
%exchanging coordinates. In order to emphasize the role of $\pu$ and $\qu$ variables we
%denote them as subscripts.
%---------------------------
%---------------------------
From this analysis we conclude that the self-energy can be written as
%===========================================================
%
%===========================================================
\bea \Sigma_c^{<}(1,2) = i\sum_{N=1}^{\infty} \frac{1}{(N+1)!N!}
\sum_{\pu\qu}\tilde{S}_{N,\pu\qu}(2) \tilde{S}_{N,\pu\qu}^*(1) .\;\;\;\;\;
 \label{s<4}
\eea
%===========================================================
%
%===========================================================
Equation \eqref{s<4} is the Lehmann-like representation of the self-energy and the main
result of this section. The irreducible part of products in Eq.~\eqref{sigma_lesser} has
been transformed into products of irreducible parts in Eq.~\eqref{s<4}.  The product of a
$\tilde{S}$-diagram and a $\tilde{S}^{\ast}$-diagram yields an
irreducible self-energy diagram in which the internal times are either integrated over the
minus-branch or over the plus-branch. We call this product a {\em partition} of the
self-energy diagram.  It is now easy to show that the Fourier transform of $-i\Sigma_{c}^{<}$ has
the PSD property. Fourier transforming $\tilde{S}$ and $\tilde{S}^{\ast}$ and omitting the
dependence on the position-spin variables we find
%===========================================================
%
%===========================================================
\bea -i\Sigma_c^{<}(1,2) &=&
\sum_{N=1}^{\infty} \frac{1}{(N+1)!N!}  \int \frac{d\w}{2\pi} \frac{d\w'}{2\pi} \nn\\ &&
e^{-i\w t_{2}+i\w' t_{1} } \sum_{\pu\qu} \tilde{S}_{N,\pu\qu}(\w)
\tilde{S}_{N,\pu\qu}^*(\w') .\quad\quad
\label{ft}
\eea
%===========================================================
%
%===========================================================
In this equation the Fourier transform of $\tilde{S}(t_{2})$ (and similarly of
$\tilde{S}^{\ast}(t_{1})$) is performed over the time argument $t_{2}$ and not over the
time difference $t_2-\tau$, which is ill-defined for $\tau\rightarrow\infty$. For the
right hand side to depend only on $t_{1}-t_{2}$ the following property
%===========================================================
%
%===========================================================
%\bea 
\[
\sum_{N=1}^{\infty} \frac{1}{(N+1)!N!}  \sum_{\pu\qu}
\tilde{S}_{N,\pu\qu}(\w) 
\tilde{S}_{N,\pu\qu}^*(\w')=\mF(\w) \delta(\w-\w')
\]
%\nn\\
%\label{prop}
%\eea
%===========================================================
%
%===========================================================
has to be fulfilled. From this property we see that $\mF(\w) \geq 0 $ and inserting it
back into equation \eqref{ft} we see that $\mF(-\w)$ is the Fourier transform of the
function $-i\Sigma_c^{<}(1,2)$ with respect to the time difference 
$t_{1}-t_{2}$. 

So far we have restricted ourselves to the exact self-energy.  In
the following section we will explain how a given approximate diagrammatic expression for
the self-energy can be extended, if necessary, to a similar form as in Eq.~\eqref{s<4} by
an appropriate selection of additional half-diagrams, thereby ensuring the positivity
of its spectral function.

%Even though we worked at the zero temperature and represented $S$-diagrams in terms of
%non-interacting Green's functions these assumptions are not essential and can be easily
%generalized to interacting Green's functions and to non-zero temperatures. This will be a
%topic of forthcoming paper.

%*************************************************************
\subsection{Diagrammatic theory of positive spectral functions}
\label{diagtheory}
\begin{figure}[t]
\centering
 \includegraphics[]{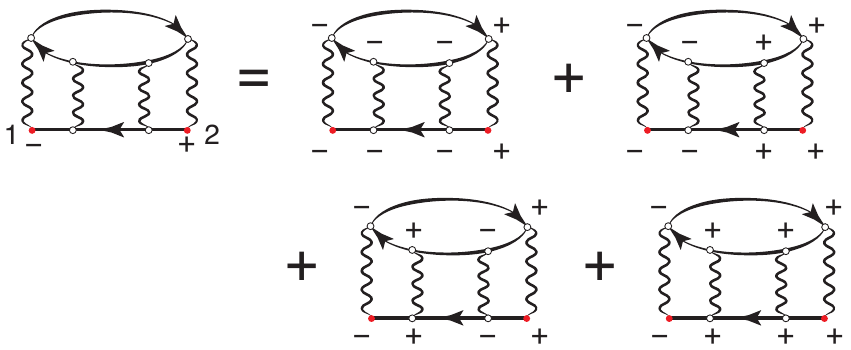}
\caption{(Color online) An example of the distribution of $+$ and $-$ 
labels over the
internal vertices of a lesser fourth order self-energy diagram.  
Divisions with 
lines $v^{+-}$ or $v^{-+}$ vanish due to the time-locality of the
interaction and therefore are not shown. The third term after the equality sign must be discarded because it contains  
an isolated island of plus signs upon cutting the $+/-$ $g$-lines (see explanation in the text).
\label{fig:diagram1}}
\end{figure}

%*************************************************************
The Lehmann-like representation of Eq.~(\ref{s<4}) brings to light a general and simple
rule to calculate the lesser component of a self-energy diagram. A diagram for
$\Si^{<}_{c}(1,2)$ has two external vertices, one with time $t_{1}$ on the
minus-branch and the other with time $t_{2}$ on the plus-branch, and a certain number of
internal vertices with times to be integrated over the Keldysh contour. If we assign to
each internal vertex a $-$ or a $+$ sign to signify that the corresponding time is
integrated over the minus or plus branch then we obtain a division of the original
self-energy diagram, and the full self-energy diagram is the sum of all of them. Since
the two-particle interaction is local in time, i.e.,
\[
v(\bx_1 t_1,\bx_2 t_2)=v(\bx_1,\bx_2)\delta(t_2-t_1),
\]
it only connects two vertices with times on the same branch of the Keldysh contour. Since
the external vertices are fixed there are $2^{n-2}$ divisions ($n>1$) for a
diagram with $n$ interaction lines.  However, not all such divisions 
contribute. As shown in Fig.~\ref{fig:ss} and in Eq.~(\ref{s<4}) 
the only divisions appearing in the expansion of the self-energy
are those in which one side of the diagram only
contains ``$-$'' vertices and the other side of the diagram only contains
``$+$'' vertices.  All other divisions must therefore be discarded.  These are 
the divisions for which we get a
piece disconnected from the external vertices 1 and 2 upon cutting 
the $+/-$ $g$-lines (hence these divisions cannot be written as the product 
of a $\tilde{S}$-$\tilde{S}^{\ast}$ diagram). The 
number of contributing
divisions clearly depends on the topological structure of the diagram. In Section
\ref{possemidef} we called such divisions partitions.
As an example consider the $\Sigma_{c}^{<}$ diagram shown in Fig.~\ref{fig:diagram1}.  The
third division vanishes since we get a piece disconnected from 1 and 2 upon cutting the
$+/-$ $g$-lines. This simple diagrammatic rule to extract the lesser self-energy can be
viewed as a generalization of the Langreth rules~\cite{devreese_linear_1976} to diagrams
which are neither a product nor a convolution of Green's functions.~\cite{ness_gw_2011}
The same diagrammatic rule can alternatively be derived by working in frequency space and
by taking into account the conservation of energy at each
vertex.\cite{veltman_unitarity_1963,kobes_discontinuities_1985}

%---------------------------
%---------------------------

There remains one issue to address before introducing our diagrammatic theory of PSD
spectral functions: How many $\tilde{S}$-$\tilde{S}^{\ast}$ diagrams do lead to the same
partition of a self-energy diagram?  To answer this question we need to investigate the
expansion of $\tilde{S}$ in terms of Feynman diagrams.  Due to the anti-commutation rules
for the creation and annihilation operators it follows from the definition of
$S_{N,\pu\qu}$ in Eq.~(\ref{Spq_def}) that a permutation $P$ of the labels $\pu$ and a
permutation $Q$ of the labels $\qu$ simply changes the sign of 
$S_{N,\pu\qu}$, and hence
also $\tilde{S}_{N,\pu\qu}$, by a factor $(-1)^{P+Q}$.  Therefore if we let
$\{D_{\pu\qu}^{(j)}\}$ with $j \in I_N$ be the set of all topologically inequivalent
diagrams for $\tilde{S}_{N,\pu\qu}$ that differ by more than a permutation of the $\pu$ or
$\qu$ labels, then we can write
%===========================================================
%
%===========================================================
\bea
\label{s_perm}
\tilde{S}_{N,\pu\qu} =  \sum_{j \in I_N} \sum_ {\substack{P \in \pi_{N+1} \\ 
Q \in \pi_N}} (-)^{P+Q}D_ {P(\pu)Q(\qu)}^{(j)},
\eea
%===========================================================
%
%===========================================================
where $P$ and $Q$ run over all permutations $\pi_{N+1}$ and $\pi_N$ of $N+1$ and $N$
indices respectively.
Inserting Eq.~\eqref{s_perm} back into
Eq.~\eqref{s<4} we find
%===========================================================
%
%===========================================================
\bea
\label{sigma_lesser_perms}
\Sigma_c^{<}(1,2) &=& i\sum_{N=1}^{\infty} \sum_{j_1, j_2 \in I_N} \sum_{\substack{P_1,P_2
    \in \pi_{N+1} \\ Q_1, Q_2 \in \pi_N} } \frac{(-)^{P_{1}+Q_{1}+P_{2}+Q_{2}}}{(N+1)!N!}
\nn\\ &\times& \sum_{\pu\qu}
D^{(j_2)}_{P_2(\pu)Q_2(\qu)}(2)D^{(j_1)^*}_{P_1(\pu)Q_1(\qu)}(1).  
\eea 
%===========================================================
%
%===========================================================
This expression can be simplified further by noticing that the composite permutations
$P\circ P_{i}$ and $Q\circ Q_{i}$ with $i=1,2$ yield the same contribution as the
permutations $P_{i}$ and $Q_{i}$, i.e.,
%===========================================================
%
%===========================================================
\begin{multline}
D^{(j_2)}_{P\circ P_2(\pu)Q\circ Q_2(\qu)}(2)D^{(j_1)^*}_{P\circ 
P_1(\pu)Q\circ Q_1(\qu)}(1)\\=
D^{(j_2)}_{P_2(\pu)Q_2(\qu)}(2)D^{(j_1)^*}_{P_1(\pu)Q_1(\qu)}(1),
\label{groupid}
\end{multline}
%===========================================================
%
%===========================================================
since the effect of $P$ and $Q$ is equivalent to a relabeling of the $\pu$ and
$\qu$. There are $N! (N+1)!$ such relabelings and they all give the same contribution.  We
can therefore simplify Eq.~(\ref{sigma_lesser_perms}) to
%===========================================================
%
%===========================================================
\begin{multline}
\label{sigma_lesser2}
\Sigma_c^{<}(1,2) =
 i\sum_{N=1}^{\infty} 
\sum_{j_1, j_2 \in I_N}\sum_{\substack{P_1 \in \pi_{N+1} \\ Q_1\in 
\pi_N} }(-)^{P_{1}+Q_{1}}
\\\times\sum_{\pu\qu}
D^{(j_2)}_{\pu\qu}(2)D^{(j_1)^*}_{P_1(\pu)Q_1(\qu)}(1).
\end{multline}
%===========================================================
%
%===========================================================
Every term of the form 
$\sum_{\pu\qu}D^{(j_2)}_{\pu\qu}(2)D^{(j_1)^*}_{P_1(\pu)Q_1(\qu)}(1)$ in
Eq. (\ref{sigma_lesser2}) corresponds to a unique partition of a $\Si^{<}_{c}$
diagram. Vice versa, every partition of a $\Si^{<}_{c}$ diagram can be written as the
product of a unique $D$-$D^{\ast}$ diagram for otherwise there should exist more than one
way to cut the self-energy along the $+/-$ $g$-lines.

%---------------------------
\begin{figure}[t]
\centering
       \includegraphics[width=0.9\columnwidth]{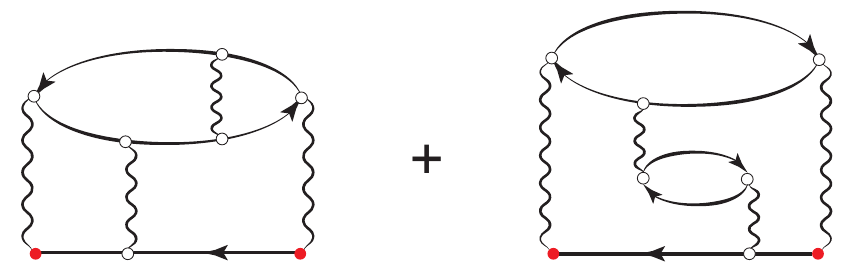}
\caption{(Color online) An approximate self-energy of many-body perturbation theory.}
 \label{exotic}
\end{figure}
\begin{figure}[t]
\centering
       \includegraphics[]{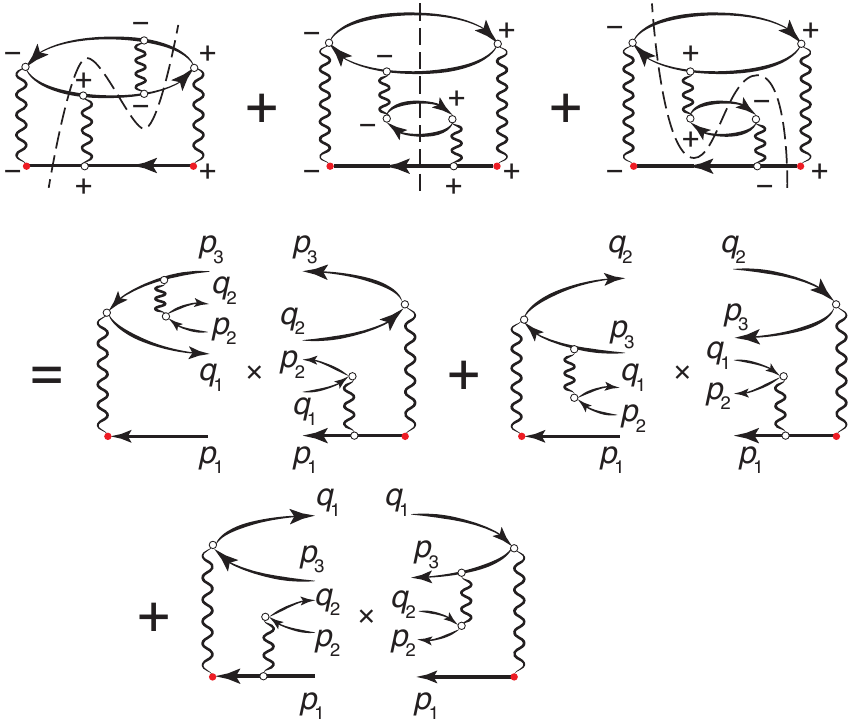}
\caption{(Color online) Partitions of the self-energy diagrams of Fig.~\ref{exotic} with
  five $+/-$ $g$-lines and decomposition in terms of $D$-$D^{\ast}$ diagrams.}
 \label{exotic-partition}
\end{figure}

%---------------------------
Equation~(\ref{sigma_lesser2}) is an exact rewriting of $\Si^{<}_{c}$ in terms of $D$
diagrams. MBPT approximations to the self-energy
consist of the sum of a (finite or infinite) subset of diagrams. An exotic approximation
could be, e.g., the one of Fig.~\ref{exotic}. Each diagram contains four interaction lines
($n=4$) and, therefore, it is divided in $2^{n-2}=4$ different ways. It is a simple
exercise to see that the left diagram is the sum of three partitions with three 
$+/-$ $g$-lines and one partition with five $+/-$ $g$-lines whereas the right
diagram is the sum of two partitions with three $+/-$ $g$-lines and two partitions with
five $+/-$ $g$-lines. In Fig.~\ref{exotic-partition} we
display, e.g., all partitions with five $+/-$ $g$-lines ($N=2$ in
Eq.~(\ref{sigma_lesser2})) and how to write them as $D$-$D^{\ast}$ diagrams. There are two
different $D$ diagrams, say $D^{(a)}$ and $D^{(b)}$, which are glued as
%===========================================================
%
%===========================================================
\bea
\sum_{\pu\qu}\left[D^{(a)}_{p_{1}p_{2}p_{3}q_{1}q_{2}} (2) \left(D^{(b)^{\ast}}_{p_{1}p_{2}p_{3}q_{1}q_{2}} (1)
  -D^{(b)^{\ast}}_{p_{1}p_{2}p_{3}q_{2}q_{1}} (1)\right)\right.\nn\\ 
\left.-D^{(b)}_{p_{1}p_{2}p_{3}q_{1}q_{2}} (2)D^{(a)^{\ast}}_{p_{1}p_{2}p_{3}q_{2}q_{1}} (1)
  \right].  \nn
\eea 
%===========================================================
%
%===========================================================
The products of the half-diagrams is 
represented  in the same order as they 
appear in this mathematical expression in Fig.~\ref{exotic-partition} after the equality 
sign.  The diagrams
$D^{(b)}$ in the first line differ only in a permutation of the labels
$q_1$ and $q_2$ as shown in the left half-diagrams of Fig.~\ref{exotic-partition}. Furthermore, the
right half-diagram of the first term ($D^{(a)}$) and the left half-diagram 
of the last term ($D^{(a)^{\ast}}$) have 
the same topological structure but differ in the labeling of $q_1$ and $q_2$.

This example is instructive since it highlights the general structure of a MBPT
approximation to the self-energy.  The partitioning
leads to an expression of the form of Eq.~(\ref{sigma_lesser2}) where 
the domains of the summation indices and the set of permutations are 
restricted.  The couple $(j_{1},j_{2})$ runs over a subset
$\mathcal{I}_{N}\subset I_{N}\times I_{N}$ of the product set of the 
topologically inequivalent
half-diagrams. In the example of Fig.~\ref{exotic-partition} we have
$\mathcal{I}_{2}=\{(a,b),(b,a)\}$. 
Given the couple $(j_{1},j_{2})\in \mathcal{I}_{N}$ the permutations $P_{1}$ and
$Q_{1}$ run over a subset $\pi_{N+1,p}^{(j_{1}j_{2})}\subset \pi_{N+1}$ and
$\pi_{N,q}^{(j_{1}j_{2})}\subset \pi_{N}$ of the permutation groups 
$\pi_{N+1}$ and $\pi_{N}$.  In the
example of Fig.~\ref{exotic-partition} for the couple  $(b,a)$ (the first two
terms after the equality sign) we have the subsets $\pi_{3,p}^{(ba)}=\{
\boldmath{1} \}$ and $\pi_{2,q}^{(ba)}=\{ \boldmath{1}, Q\}$ with 
$Q(q_{1},q_{2})=(q_{2},q_{1})$, whereas for the couple 
$(a,b)$ (the last term after the equality sign) we have the subsets
$\pi_{3,p}^{(ab)}=\{1\}$
and $\pi_{2,q}^{(ab)}=\{Q\}$. In Fig.~\ref{exotic-partition} we
considered in detail the $D$-$D^{\ast}$ diagrams 
of the self-energy of Fig.~\ref{exotic} belonging to the set 
$\mathcal{I}_2$. The set $\mathcal{I}_1$ in
which we cut three $+/-$ $g$-lines can be analyzed similarly. All other sets
$\mathcal{I}_N$ with $N>2$ are empty in this case.  In general, 
however, we have for an approximate MBPT self-energy
%===========================================================
%
%===========================================================
\bea
\label{sigma_mbpt}
\Sigma_{c}^{<}(1,2) 
&=& i\sum_{N=1}^{\infty}
\sum_{(j_1, j_2) \in \mathcal{I}_N} \sum_{\substack{P_1\in 
\pi^{(j_{1}j_{2})}_{N+1,p} \\ Q_1 \in \pi^{(j_{1}j_{2})}_{N,q}} }(-)^{P_{1}+Q_{1}}
\nn\\
&\times& \sum_{\pu\qu}
D^{(j_2)}_{\pu\qu}(2)D^{(j_1)^*}_{P_1(\pu)Q_1(\qu)}(1),
\eea
%===========================================================
%
%===========================================================
where the sets $\mathcal{I}_N$ may contain zero, a finite or an infinite number of elements.

An important remark to be made regarding Eq.~(\ref{sigma_mbpt}) is that it does not, in general,
fulfill the PSD property.  
%---------------------------
%---------------------------
Our analysis shows, however, how to formulate the diagrammatic rules to transform a MBPT
self-energy into a PSD self-energy by adding the minimal number of partitions. For the PSD
property to be fulfilled the self-energy should have the structure of Eq.  (\ref{s<4})
with some approximate $\tilde{S}_{N,\pu\qu}$. Let $\tilde{I}_N \subset I_N$ be the
smallest subset such that
%===========================================================
%
%===========================================================
\be \tilde{I}_N \times \tilde{I}_N\supset \mathcal{I}_{N},
\label{rule1}
\ee
%===========================================================
%
%===========================================================
and $\tilde{\pi}_{N+1,p}$ and $\tilde{\pi}_{N,q}$ be the smallest {\em subgroups} of the
permutation groups $\pi_{N+1}$ and $\pi_{N}$ with the property
%===========================================================
%
%===========================================================
\bea
\tilde{\pi}_{N+1,p}\supset \bigcup_{(j_1,j_2)\in \mathcal{I}_N} 
\pi^{(j_1 j_2)}_{N+1,p}\;,\label{rule2}
\\
\tilde{\pi}_{N,q}\supset\bigcup_{(j_1,j_2)\in \mathcal{I}_N} 
\pi^{(j_1 j_2)}_{N,q}.
\label{rule3}
\eea
%===========================================================
%
%===========================================================
Mathemetically $\tilde{\pi}_{N+1,p}$ and $\tilde{\pi}_{N,q}$ are given by the \emph{intersection} of all
subgroups containing the subsets $\pi^{(j_1 j_2)}_{N+1,p}$ and $\pi^{(j_1 
j_2)}_{N,q}$. The self-energy
%===========================================================
%
%===========================================================
\bea
\label{sigma_psd1}
\Sigma_{c,\rm PSD}^{<}(1,2) 
&=& i\sum_{N=1}^{\infty}
\sum_{j_1, j_2 \in\tilde{I}_N} \sum_{\substack{P_1\in 
\tilde{\pi}_{N+1,p} \\ Q_1 \in \tilde{\pi}_{N,q}} }(-)^{P_{1}+Q_{1}}
\nn\\
&\times& \sum_{\pu\qu}
D^{(j_2)}_{\pu\qu}(2)D^{(j_1)^*}_{P_1(\pu)Q_1(\qu)}(1)
\eea
%===========================================================
%
%===========================================================
contains all partitions of Eq. (\ref{sigma_mbpt}) plus other partitions, and each
partition is counted only once. Furthermore, taking into account that
$\tilde{\pi}_{N+1,p}$ and $\tilde{\pi}_{N,q}$ are two subgroups, 
Eq.~(\ref{groupid})
is valid for all $P\in \tilde{\pi}_{N+1,p}$ and $Q\in \tilde{\pi}_{N,q}$. Hence we can
rewrite Eq.~(\ref{sigma_psd1}) to bring out its product structure. We have
%===========================================================
%
%===========================================================
\bea
\label{sigma_psd2}
\Sigma_{c,,\rm PSD}^{<}(1,2) 
&=& i\sum_{N=1}^{\infty}
\sum_{j_1, j_2 \in \tilde{I}_N} 
\sum_{\substack{P_1,P_2 \in\tilde{\pi}_{N+1,p} \\ Q_1, Q_2 \in \tilde{\pi}_{N,q}} }
\!\!\!\!\!\! \frac{(-)^{P_{1}+Q_{1}+P_{2}+Q_{2}}}{d_{N+1,p}d_{N,q}} 
\nn\\
&\times& 
\sum_{\pu\qu}
D^{(j_2)}_{P_2(\pu)Q_2(\qu)}(2)D^{(j_1)^*}_{P_1(\pu)Q_1(\qu)}(1),
\eea
%===========================================================
%
%===========================================================
where $d_{N+1,p}$ and $d_{N,q}$ are the dimensions of the subgroups $\tilde{\pi}_{N+1,p}$ and
$\tilde{\pi}_{N,q}$ respectively. The self-energy in Eq.~(\ref{sigma_psd2}) is clearly PSD. 
It is also clear that any
reduction of the sets $\tilde{I}_N$, $\tilde{\pi}_{N+1,p}$ and $\tilde{\pi}_{N,q}$ would either
not include the original MBPT diagrams or would not fulfill the PSD property. Thus
Eq. (\ref{sigma_psd2}) contains the minimal number of partitions of self-energy diagrams
to correct the MBPT self-energy.

This concludes the diagrammatic theory to generate PSD spectral functions. In the next
sections we work out explicitly some text-book examples and derive the leading PSD
self-energy diagrams with {\em vertex corrections} and {\em screened interaction}, thus
going beyond the GW approximation.

%*************************************************************
\subsection{Self-consistency}
\label{selfconsistency}
%*************************************************************

Before we discuss some examples in detail we would first like to 
address the issue of
self-consistency which plays an important role in so-called conserving
approximations.\cite{stefanucci_nonequilibrium_2013,baym_self-consistent_1962} So far we used the 
noninteracting Green's functions $g$ to evaluate the
diagrams. Suppose that a specific selection of partitions
guarantees the positivity of the spectral function for 
$\Sigma_{c,\rm PSD}^\lessgtr [g]$,
where we indicate explicitly the functional dependence of the self-energy on $g$.  Then
we can use this self-energy in the Dyson equation to evaluate a new 
Green's function, which
we may call $G$ to distinguish it from the noninteracting $g$.  In the
next step we can evaluate our approximate diagrammatic expression for the self-energy in
terms of $G$, i.e., we evaluate $\Sigma_{c,\rm PSD}^\lessgtr [G]$.  The natural question to
ask then is whether $\Sigma_{c,\rm PSD}^\lessgtr [G]$ still has the PSD
property. We now demonstrate that this is indeed the case
by a modification of the derivation in Sections \ref{possemidef} and \ref{diagtheory}.

The largest modification involves relations (\ref{idemp_1}) and 
(\ref{idemp_2}) since they are
not valid anymore for general dressed Green's functions. This is, for example, easily
demonstrated for the exact interacting lesser Green's function of an 
$N$-particle system with ground state energy $E_{0}$,
which has the Lehmann representation
%===========================================================
%
%===========================================================
\be
G^<_{\bx\bx'} (t, t') = i \sum_\a e^{-i \Omega_\a (t-t')}   f_\a 
(\bx)   f_\a^* (\bx' ).
\label{lrexg}
\ee
%===========================================================
%
%===========================================================
Here $\a$ labels the many-body eigenstates $\Psi_{\a,N-1}$  with energy  
$E_{\a,N-1}$ of the system with $N-1$ 
particles,
$\Omega_\a = E_{\a,N-1}- E_{0,N}$ are the removal energies and $f_\a 
(\bx) = \langle \Psi_{\a,N-1}
| \hat{\psi} (\bx) | \Psi_0 \rangle$ the so-called Dyson orbitals. 
Although the Lehmann representation (\ref{lrexg})
looks formally identical to the expansion in Eq.~(\ref{gless_exp}) 
for $g$, it does not allow us to derive Eq.~(\ref{idemp_1}) anymore 
since the nonvanishing $f_\a (\bx)$ form an overcomplete and 
nonorthonormal one-particle basis-set (in a noninteracting system most $f_\a (\bx)$ 
are zero and the nonvanishing ones form an orthonormal basis-set). Our strategy, therefore,
is to replace Eqs. (\ref{gless_exp}) and (\ref{ggrt_exp}) with a different
relation that still allows us to formulate  a cutting rule.  Crucial in our
reasoning is that a PSD self-energy generates
a PSD spectral function for $G$, as was explained just below
Eq.~(\ref{GRA}). This implies that $G^<$ has the form
%===========================================================
%
%===========================================================
\be
G_{\bx\bx'}^< (t,t') = i \int \frac{d\omega}{2 \pi} \, A^{<}_{\bx\bx'} (\omega) \, e^{- i  \omega (t-t')} \nn
\ee
%===========================================================
%
%===========================================================
where the removal-part of the spectral function $A^{<}_{\bx\bx'} 
(\omega)\equiv f(\w)A_{\bx\bx'}(\w)$ is a self-adjoint 
and PSD matrix in the one-particle indices for every $\w$.
Denoting by $\bra\bx|a_{i}(\w)\ket$ the eigenstates of $A^{<}_{\bx\bx'} 
(\omega)$ with eigenvalue $a_{i}(\w)$, the spectral representation 
of this matrix can be written as
%===========================================================
%
%===========================================================
\be
A_{\bx\bx'}^{<} (\omega) = \sum_{i} \bra\bx|a_{i}(\w)\ket \, a_{i}(\w) 
\langle a_{i}(\w) | \bx' \rangle. \nn
\ee
%===========================================================
%
%===========================================================
Without loss of generality we shall assume that the eigenstates 
$|a_{i}(\w)\ket$ form an orthonormal basis-set for every $\w$.
Due to the PSD property of $A^{<}$ the eigenvalues $a_{i}\geq 0$, and therefore we 
can  define the square root of the spectral function according to
%===========================================================
%
%===========================================================
\be
\sqrt{A^{<}}_{\,\bx\bx'} (\omega) = \sum_{i} \langle \bx | a_{i} \rangle \, 
\sqrt{a_{i}}\; 
\langle a_{i} | \bx' \rangle \nn
\ee
%===========================================================
%
%===========================================================
where for notational convenience we suppressed the $\omega$-dependence of the
eigenvalues and eigenvectors.
Correspondingly we can define the square root of the lesser Green's 
function according to
%===========================================================
%
%===========================================================
\be
\sqrt{G^{ <}}_{\,\bx\bx'} (t,t') = i \int \frac{d\omega}{2 \pi} 
\,\sqrt{A^{<}}_{\,\bx\bx'} (\omega) \, 
e^{- i  \omega (t-t')} . \nn
\ee
%===========================================================
%
%===========================================================
This function has the property that
%===========================================================
%
%===========================================================
\be
i G^{<}_{\bx\bx'} (t,t') =  \int \!d\by d\bar{t} \;
\sqrt{G^{<}}_{\;\bx \by} (t,\bar{t}) 
\sqrt{G^{<}}_{\by\bx'} (\bar{t},t') \label{Gbar1}
\ee
%===========================================================
%
%===========================================================
as follows from a quick calculation using the definitions above.  Similarly $G^>$ 
is the integral over a positive spectral function, and we can 
therefore define the square root $\sqrt{G^{>}}$ with the property that
%===========================================================
%
%===========================================================
\be
-i G^>_{\bx\bx'} (t,t') =  \int \!d\by d\bar{t} \; 
\sqrt{G^{>}}_{\bx \by} (t,\bar{t}) \sqrt{G^{>}}_{\by \bx'} (\bar{t},t') .
\label{Gbar2}
\ee 
%===========================================================
%
%===========================================================
The relations (\ref{Gbar1}) and (\ref{Gbar2}) provide a new cutting rule for a
self-energy diagram with a dressed Green's function. Whenever we cut a self-energy diagram
we obtain half-diagrams with outgoing lines $\sqrt{G^{ <}}$ and $\sqrt{G^{>}}$.  Using these
modified half-diagrams we obtain an equation that is identical in structure to
Eq.~(\ref{sigma_psd1}).  The only thing that changes in 
Eq.~(\ref{sigma_psd1}) is that the sums over $\pu\qu$ are replaced by 
integrals over $\by$'s and $\bar{t}$'s.  However this
does not change the quadratic structure of the equation. We therefore conclude that 
$\Sigma_{c,\rm PSD}^< [G]$ also is PSD. From
this new self-energy we can use the Dyson equation to calculate a new Green's function 
which has again a PSD spectral function and can be decomposed as
in Eqs.~(\ref{Gbar1}) and (\ref{Gbar2}), thereby yielding yet another PSD self-energy. By
repeating the procedure  we obtain a series of PSD Green's functions.
If this series converges to a limiting Green's function then we have solved the Dyson
equation self-consistently for our approximate $\Sigma_{c,\rm PSD}^<$ in which both
the Green's function and the self-energy are PSD.

The conclusion of this analysis is that our diagrammatic approach to 
PSD spectral functions also applies to self-consistent perturbation theory. 
Of course, in order to avoid double countings, the 
partitions of which $\Si^{<}_{c,\rm PSD}$ is made of should be skeletonic, i.e.,   
should not contain self-energy insertions.\cite{stefanucci_nonequilibrium_2013}
An important approximation that has this structure is the GW approximation which we 
study in more detail below. In general, however, it should be 
emphasized that a PSD self-energy made exclusively of all the 
partitions of conserving 
diagrams is rare. In fact, approximations which are 
simultaneously conserving and PSD are exceptional.

%**************************************************************************************
\section{Examples}
\label{examples}
%**************************************************************************************

\begin{figure}[t!]
\centering
\includegraphics[width=0.9\columnwidth]{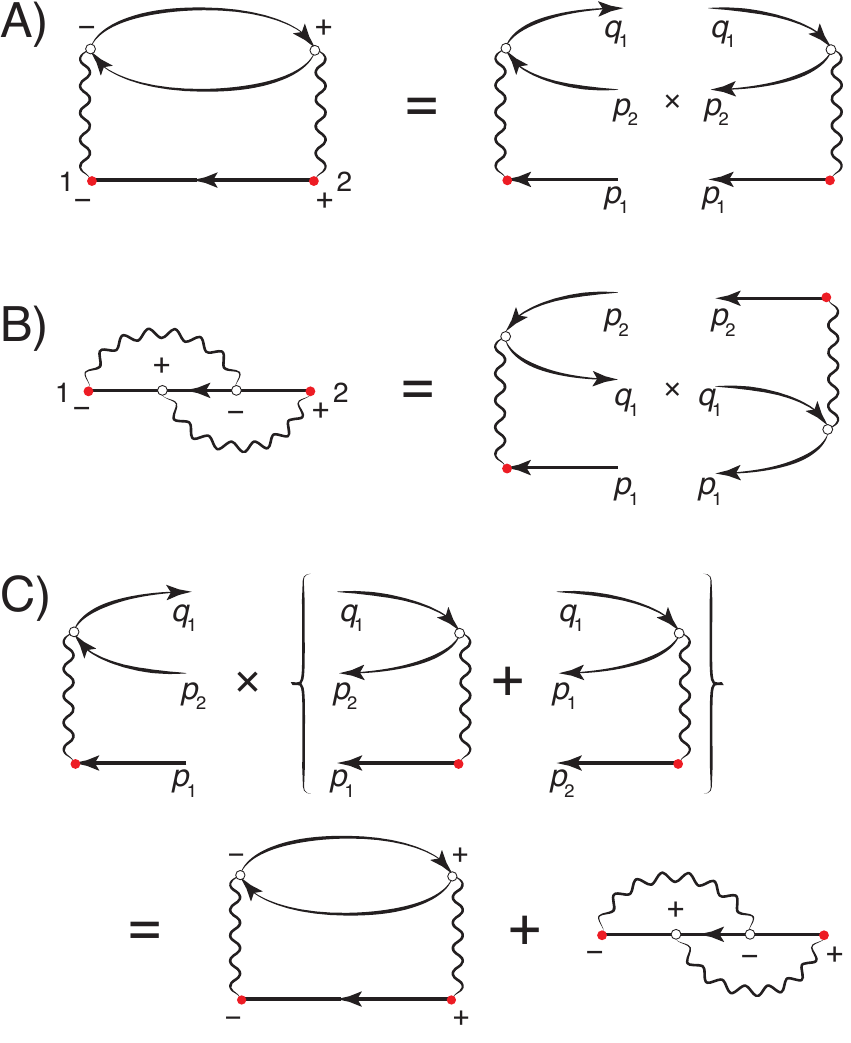}
\caption{(Color online) (A) Partition of the
  first bubble diagram and the resulting half-diagrams.  (B) 
  Partition of the second-order exchange diagram and the resulting
  half-diagrams.  (C) In order to form a a complete square with the 
  second-order exchange diagram we need to add the right half-diagram with 
  permuted  $p_1$ and $p_2$ labels. This yields the 2B approximation.
\label{fig:example1}}
\end{figure}

In this section we apply the formalism developed in Section 
\ref{sec:theory} to some illustrative
examples. 
\paragraph{Single bubble diagram}
Let us first consider the first bubble diagram shown in Fig.~\ref{fig:example1}(A).  The
lesser component of this self-energy reads
%===========================================================
%
%===========================================================
\bea
\Si_c^{<}(1,2) &=& \int d\bx_3 \int d\bx_4 v(\bx_1,\bx_3)
g^{<}_{\bx_1\bx_2}(t_1,t_2)\nn\\
&\times& g^{<}_{\bx_3\bx_4}(t_1,t_2)  g^{>}_{\bx_4\bx_3}(t_2,t_1)v(\bx_2,\bx_4)
\eea 
%===========================================================
%
%===========================================================
which can be partitioned in only one way. Upon cutting along the $+ / -$ $g$-lines we find
%===========================================================
%
%===========================================================
\be
\label{bubbe_in_half}
\Si^{<}_{c}(1,2) =  i\sum_{\pu\qu}
D_{p_1p_2q_1}^{(a)} (2) 
D^{(a)^*}_{p_1p_2q_1}(1),
\ee
%===========================================================
%
%===========================================================
where the $D$ diagram reads 
%===========================================================
%
%===========================================================
\bea
D^{(a)^*}_{p_1p_2q_1}(1)&=& \int d\bx_3 v(\bx_1,\bx_3)
g^{<}_{\bx_1\,p_1}(t_1,\t) \nn\\
&\times&g^{<}_{\bx_3\,p_2}(t_{1},\t)
g^{>}_{q_1\,\bx_3}(\t,t_1) .
\label{dbubble}
\eea
%===========================================================
%
%===========================================================
Equation~\eqref{bubbe_in_half} is already of the form in Eq.~\eqref{sigma_psd1} and
therefore the first bubble diagram produces a PSD spectrum.
%---------------------------
%---------------------------
\paragraph{Second-order exchange}
The exchange diagram of the first bubble diagram is shown in Fig.~\ref{fig:example1}(B)
and the application of the formalism is more interesting. The lesser component of the
self-energy now reads
%===========================================================
%
%===========================================================
\bea
\Si_c^{<}(1,2) &=&-\int d\bx_3 \int d\bx_4 v(\bx_1,\bx_3)
g^{<}_{\bx_1\bx_4}(t_1,t_2)\nn\\
&\times& g^{>}_{\bx_4\bx_3}(t_2,t_1) g^{<}_{\bx_3\bx_2}(t_1,t_2) v(\bx_4,\bx_2). \nn
\eea
%===========================================================
%
%===========================================================
This diagram too can be partitioned in only one way.  However, upon cutting the $+ /
-$ $g$-lines we find a product of $D$ diagrams differing by a permutation
$P(p_{1},p_{2})=(p_{2},p_{1})$
%===========================================================
%
%===========================================================
\bea
\Sigma_c^{<}(1,2) =i \sum_{\pu\qu} (-)^{P}
D_{p_1p_2q_1}^{(a)} (2) D^{(a)^*}_{p_{2}p_{1}q_1}(1)\nn
\eea
%===========================================================
%
%===========================================================
where $D_{p_1p_2q_1}^{(a)}$ is the same as in Eq. (\ref{dbubble}).  The smallest subgroup
of $\pi_2$ which contains $P$ is $\pi_2$ itself and the domain $j_1=a$ and $j_2=a$ is
already of the form $\tilde{I}_1 \times \tilde{I}_1$. Therefore we can form a PSD
self-energy by taking $\tilde{I}_1=\{a\}$, $\tilde{\pi}_{2,p} = \pi_2$ and
$\tilde{\pi}_{1,q}=\{\boldmath{1}\}=\pi_1$. In this way we end up with the diagrams shown in
Fig.~\ref{fig:example1}(C). This is the second-Born (2B) approximation, which we have now shown
to give a PSD spectrum. The second-order exchange diagram is particularly instructive since it shows that the
PSD outcome of the sum of MBPT diagrams is not the sum of the PSD outcome of the MBPT
diagrams taken separately. Indeed the PSD outcome of the first-bubble diagram is the first
bubble-diagram itself whereas the PSD outcome of the second-order exchange diagram is the
2B approximation. This implies that if we had summed the PSD outcomes of these
two separate diagrams we would have counted the first-bubble diagram twice. No double
counting occurs if we apply the rules in Eqs. 
(\ref{rule1}-\ref{rule3}) to the 2B
approximation; the PSD outcome would be the 2B approximation itself.

 %---------------------------
\begin{figure}[t!]
\centering
\includegraphics[width=0.9\columnwidth]{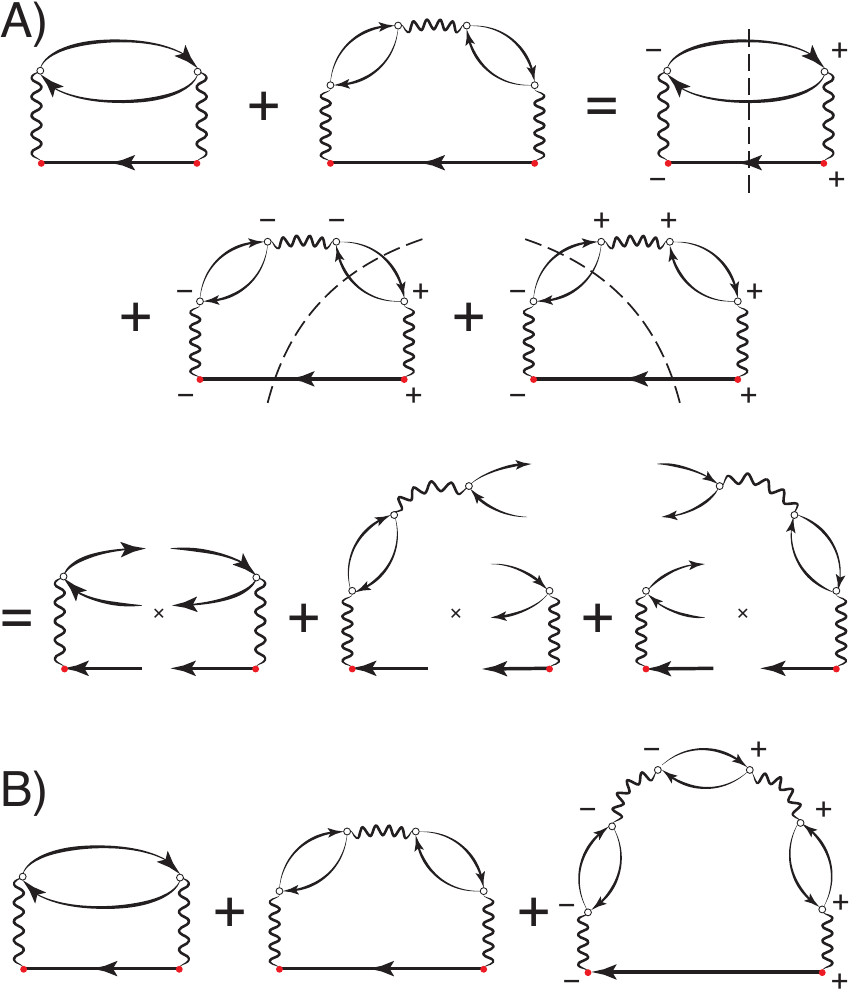}
\caption{(Color online) (A) Partitions of the
  first and second bubble diagrams and the resulting half-diagrams.  (B) The
  minimal completion of the square which yields a PSD spectrum.
\label{fig:example3}}
\end{figure}
%---------------------------

\paragraph{Two bubbles}
As a third example let us consider the sum of the first and second bubble diagrams as
shown in Fig.~\ref{fig:example3}(A).  After distributing the pluses and 
minuses over the
internal vertices we find that the lesser self-energy diagram can be written in terms of
two types of $D$-diagrams
%===========================================================
%
%===========================================================
\bea
\Sigma_c^{<}(1,2) \!\!&=&\! i\sum_{\pu\qu} \Big [
D_{p_1p_2q_1}^{(a)}(2)D_{p_1p_2q_1}^{(a)^*}(1)\nn\\
&+&\!\!D_{p_1p_2q_1}^{(b)}(2)D_{p_1p_2q_1}^{(a)^*}(1)+
D_{p_1p_2q_1}^{(a)}(2)D_{p_1p_2q_1}^{(b)^*}(1) \Big].\nn
\eea
%===========================================================
%
%===========================================================
The three $D$-$D^{\ast}$ diagrams in this expression are represented 
in the bottom line of Fig.~\ref{fig:example3}(A).
We observe that the only permutation
appearing is the identity permutation.
Therefore,
according to the rules in Eqs. (\ref{rule1}-\ref{rule3}), we only need to find the smallest
$\tilde{I}_1$ such that $\tilde{I}_1 \times \tilde{I}_1 \supset
\{(a,a),(a,b),(b,a)\}$. This is simply $\tilde{I}_1=\{a,b\}$. Thus, 
the PSD outcome of the
self-energy is
%===========================================================
%
%===========================================================
\be
\Sigma_{c,\rm PSD}^{<}(1,2) =
 \sum_{j_1j_2\in \tilde{I}_1}\sum_{\pu\qu}
D_{p_1p_2q_1}^{(j_2)}(2)D_{p_1p_2q_1}^{(j_1)^{\ast}}(1)\nn
\ee
%===========================================================
%
%===========================================================
and the corresponding diagrams are shown in Fig.~\ref{fig:example3}(B). In accordance with our
notation a diagram with no plus/minus on the internal vertices represents the full
diagram, i.e., the sum of all possible partitions.

\begin{figure}[t!]
\centering
\includegraphics[width=0.9\columnwidth]{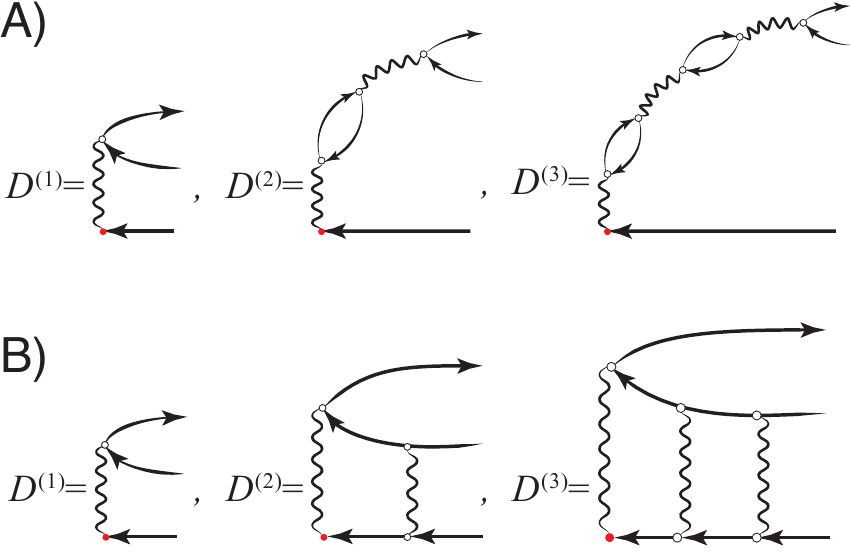}
\caption{(Color online) (A) Partitions of the bubble diagrams of the GW approximation.
  (B) Partitions of the ladder diagrams of the T-matrix approximation.
\label{fig:g0w0}}
\end{figure}

\paragraph{GW and T-matrix approximations}
With similar arguments it is easy to show that the GW self-energy has the PSD
property. We take the RPA $W=v+vPW$ with
%===========================================================
%
%===========================================================
\be
P_{\bx_{1}\bx_{2}}(z_{1},z_{2})=-ig_{\bx_{1}\bx_{2}}(z_{1},z_{2})g_{\bx_{2}\bx_{1}}(z_{2},z_{1}).
\ee Then \bea W^{<}&=&v\left(P^{<}+P^{\minus\minus}vP^{<}+P^{<}vP^{\plus\plus}\right.
\nn\\ &+&P^{\minus\minus}vP^{\minus\minus}vP^{<}+P^{\minus\minus}vP^{<}vP^{\plus\plus}+
P^{<}vP^{\plus\plus}vP^{\plus\plus} \nn\\&+&\left.\ldots\right)v ,
\eea 
%===========================================================
%
%===========================================================
where we took into account that if $P^{\plus\plus}$ appears to the left of $P^{<}$ and/or
$P^{\minus\minus}$ appears to the right of $P^{<}$ then the division 
is not a partition  since a cut
along the $+/-$ $g$-lines generates a disconnected $+$ and/or $-$ 
island, see Section \ref{diagtheory}.
It is a matter of simple algebra to prove that
%===========================================================
%
%===========================================================
\bea \Si_{\rm GW}^{<}(1,2)&\equiv&ig^{<}(1,2)W^{<}(1,2)
\nn\\&=&i \sum_{j_{1}j_{2}=1}^{\infty}\sum_{\pu\qu}
D_{p_1p_2q_1}^{(j_2)}(2)D_{p_1p_2q_1}^{(j_1)^{\ast}}(1),
\label{sigmagw}
\eea
%===========================================================
%
%===========================================================
where the diagrams $D^{(j)}$ are defined as in Fig.~\ref{fig:g0w0}(A). Equation
(\ref{sigmagw}) is clearly of the form in Eq. (\ref{sigma_psd1}).  In a similar fashion it
can be shown that the symmetrized version of the T-matrix approximation also has the PSD
property since it can be written as in the second row of Eq. (\ref{sigmagw}) with $D$
diagrams given in Fig.~\ref{fig:g0w0}(B).

%---------------------------

%**************************************************************************************
\section{PSD self-energy beyond GW}
\label{beyondGW}
%**************************************************************************************

\begin{figure}[b]
\centering
\includegraphics[width=\linewidth]{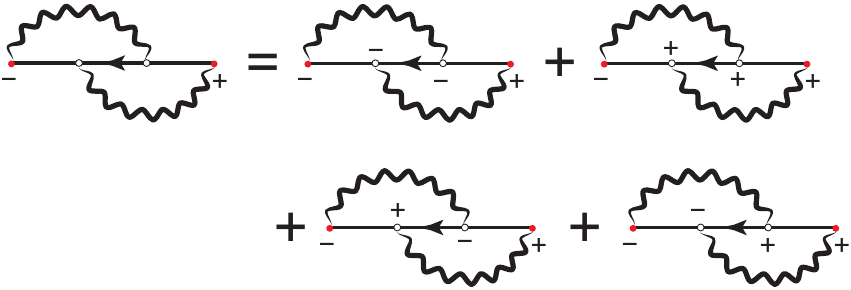}
\caption{(Color online) Next to leading order self-energy in the screened interaction
  $W$. Since $W$ is non-local in time, the thick wavy lines denoting the screened
  interaction can connect points on different branches of
  the Keldysh contour.}
\label{sigmavert}
\end{figure}
%---------------------------

The GW self-energy is the leading order approximation in the screened interaction
$W$. Despite the numerous successful applications of the GW approximation in reproducing
experimental spectra there is also a commensurable number of examples for which the GW
approximation fails, pointing out the importance of including vertex corrections. The next
to leading order approximation in $W$ is represented by the diagram $\Si_{\rm vert}$ of
Fig.~\ref{sigmavert} and contains a vertex correction. Unfortunately the self-energy
$\Si_{\rm GW+v}\equiv \Si_{\rm GW}+\Si_{\rm vert}$ is not PSD and the resulting spectral
function has the undesired feature of being negative in some region of the frequency
space. In this section we use the rules of Eqs.  (\ref{rule1}-\ref{rule3}) to identify the
minimal number of additional partitions for constructing the
leading order beyond-GW self-energy with the PSD property.

On the right hand side of Fig.~\ref{sigmavert} the self-energy $\Si_{\rm vert}$ is written
as the sum of four partitions. Let us determine the underlying $D$ diagrams. 
We observe that the screened interactions $W^{\plus\plus}$ or
$W^{\minus\minus}$ can be partitioned in only one way with all internal polarizations
$P^{\plus\plus}$ or $P^{\minus\minus}$. Indeed a partition of $W^{\plus\plus}$ in which
appears a $P^{>}$ does necessarily contain also a $P^{<}$ and hence the cut along the
$+/-$ $g$-lines would generate a disconnected island. A similar argument applies to
$W^{\minus\minus}$.
%----------------------------
\begin{figure}[t]
\centering
\includegraphics[width=\columnwidth]{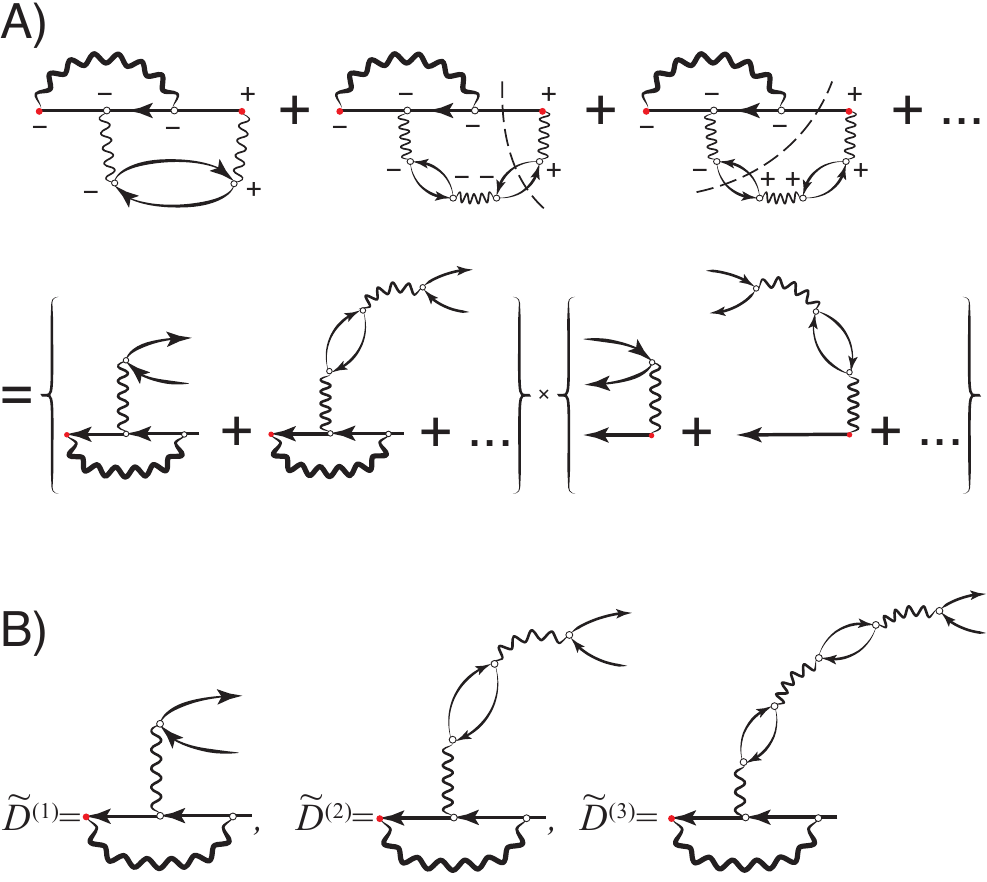}
\caption{(Color online) (A) Partitions of the first diagram on the right hand side of
  Fig. \ref{sigmavert}. (B) Definition of the
  $\tilde{D}^{(j)}$ diagrams.
 \label{fig:example4}}
\end{figure}
%----------------------------
%----------------------------
\begin{figure}[b]
\centering
\includegraphics[width=\columnwidth]{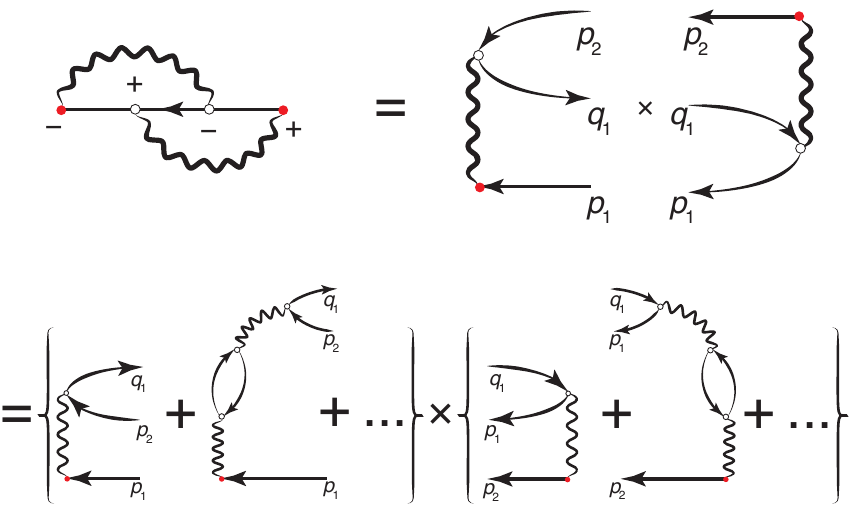}
\caption{(Color online) Partitions of the third diagram on the right hand side of
  Fig. \ref{sigmavert}.}
\label{fig:fig9}
\end{figure}
%----------------------------
Accordingly the first diagram on the right hand side of Fig.~\ref{sigmavert} is the sum of
the partitions shown in Fig. \ref{fig:example4}(A), and can be written as the sum of products
between the $D^{(j)}$ diagrams of the GW approximation, see Fig. \ref{fig:g0w0}(A), and the
$\tilde{D}^{(j)}$ diagrams displayed in Fig.  \ref{fig:example4}(B). The
same is true for the second diagram of Fig.~\ref{sigmavert} provided we exchange $D^{(j)}$
with $\tilde{D}^{(j)}$. The partitions of the third diagram of 
Fig.~\ref{sigmavert} are
very simple due to the presence of only $W^{\plus\plus}$ and $W^{\minus\minus}$, see
Fig. \ref{fig:fig9}. The result is the sum of products between $D^{(j)}$ diagrams and
permuted $D^{(j)}$ diagrams. Finally the fourth diagram of Fig.~\ref{sigmavert} is
partitioned as illustrated in Fig.~\ref{D5}(A), and can be written as the sum of products
between two $D^{(ij)}$ diagrams, see Fig.~\ref{D5}(B), where $i$ refers to
the number of top bubbles and $j$ to the number of bottom bubbles. In conclusion (omitting
the dependence on $1$ and $2$)
%----------------------------
\begin{figure}[t]
\centering
\includegraphics[width=\columnwidth]{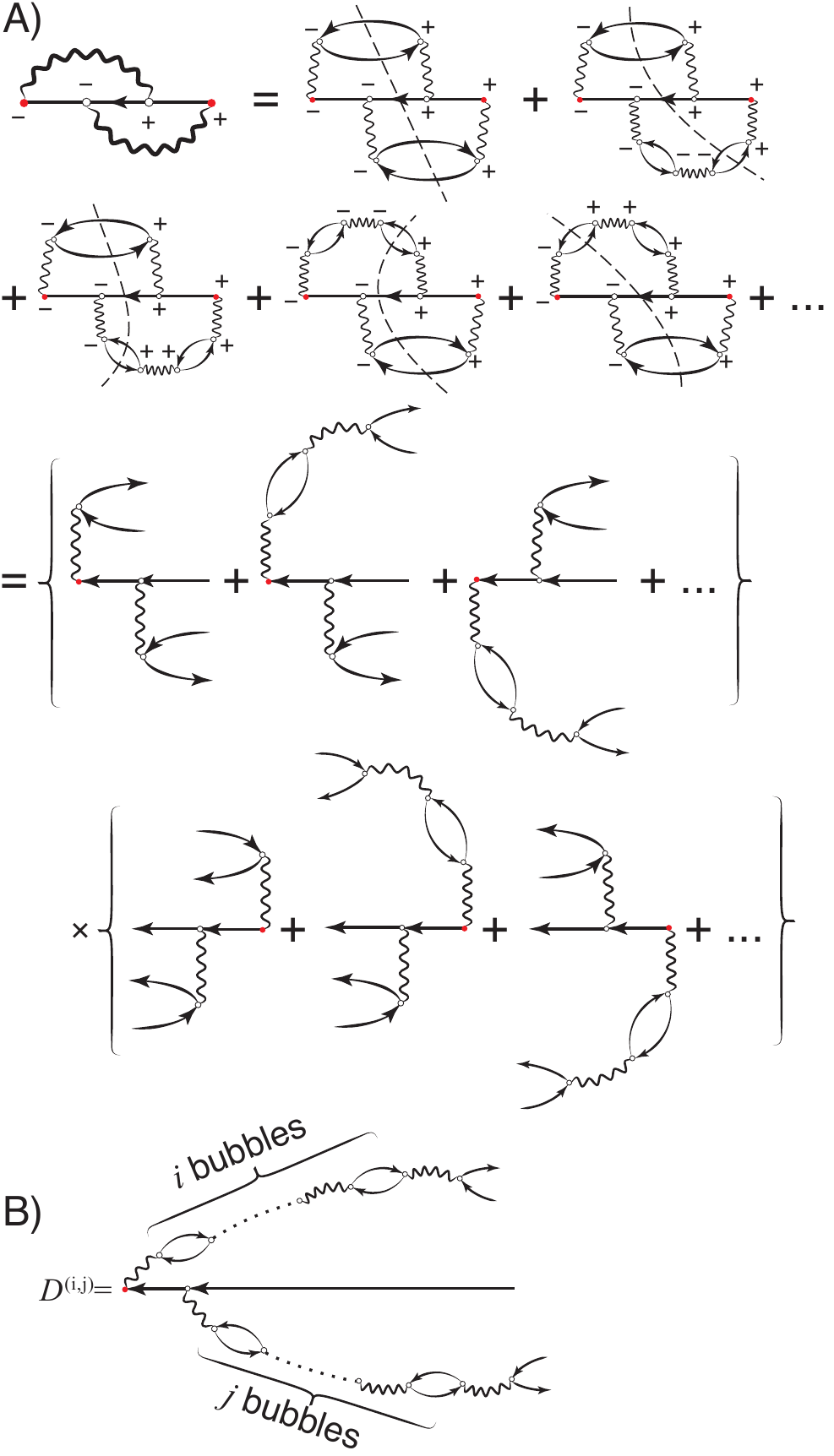}
\caption{(Color online) (A) Partitions of the fourth diagram on the right hand side of
  Fig. \ref{sigmavert}. (B) Definition of the $D^{(i,j)}$ diagrams.}
\label{D5}
\end{figure}
%----------------------------
%===========================================================
%
%===========================================================
\bea
\Si^{<}_{\rm GW+v}&=&i\sum_{j_{1}j_{2}=1}^{\infty}\sum_{\pu\qu}
\left(D^{(j_{2})}_{p_{1}p_{2}q_{1}}D^{(j_{1})^{\ast}}_{p_{1}p_{2}q_{1}}
+D^{(j_{2})}_{p_{1}p_{2}q_{1}}\tilde{D}^{(j_{1})^{\ast}}_{p_{1}p_{2}q_{1}}
\right.
\nn\\
&+&\left.
\tilde{D}^{(j_{2})}_{p_{1}p_{2}q_{1}}D^{(j_{1})^{\ast}}_{p_{1}p_{2}q_{1}}
-D^{(j_{2})}_{p_{1}p_{2}q_{1}}D^{(j_{1})^{\ast}}_{p_{2}p_{1}q_{1}}
\right)
\nn\\
&+&i\sum_{i_{1}j_{1}i_{2}j_{2}=1}^{\infty}\sum_{\pu\qu}
D^{(i_{2}j_{2})}_{p_{1}p_{2}p_{3}q_{1}q_{2}}D^{(i_{1}j_{1})^{\ast}}_{p_{1}p_{2}p_{3}q_{1}q_{2}}.
\label{sigmagwv}
\eea
%===========================================================
%
%===========================================================
Thus $\Si_{\rm GW+v}$ is the sum of a contribution $\Si_{3}$ containing partitions with
three $+/-$ $g$-lines and a contribution $\Si_{5}$ (the last term in
Eq. (\ref{sigmagwv})) containing partitions with five $+/-$ $g$-lines.  From the rules in
Eqs. (\ref{rule1}-\ref{rule3}) we see that $\Si_{5}$ has already a PSD structure. Instead
$\Si_{3}$ should be corrected according to $\Si_{3}\to \Si_{3,\rm PSD}$ with
%===========================================================
%
%===========================================================
\bea
\Si^{<}_{3,\rm PSD}&=& i\sum_{j_{1}j_{2}=1}^{\infty}\sum_{P\in\pi_{2}}
(-)^{P}\sum_{\pu\qu}
\left( 
D^{(j_{2})}_{p_{1}p_{2}q_{1}}+\tilde{D}^{(j_{2})}_{p_{1}p_{2}q_{1}}\right)
\nn\\
&\times&
\left(
D^{(j_{1})^{\ast}}_{P(p_{1})P(p_{2})q_{1}}+\tilde{D}^{(j_{1})^{\ast}}_{P(p_{1})P(p_{2})q_{1}}\right).
\eea
%===========================================================
%
%===========================================================
%----------------------------
\begin{figure}[H!]
\centering
\includegraphics[width=0.8321\columnwidth]{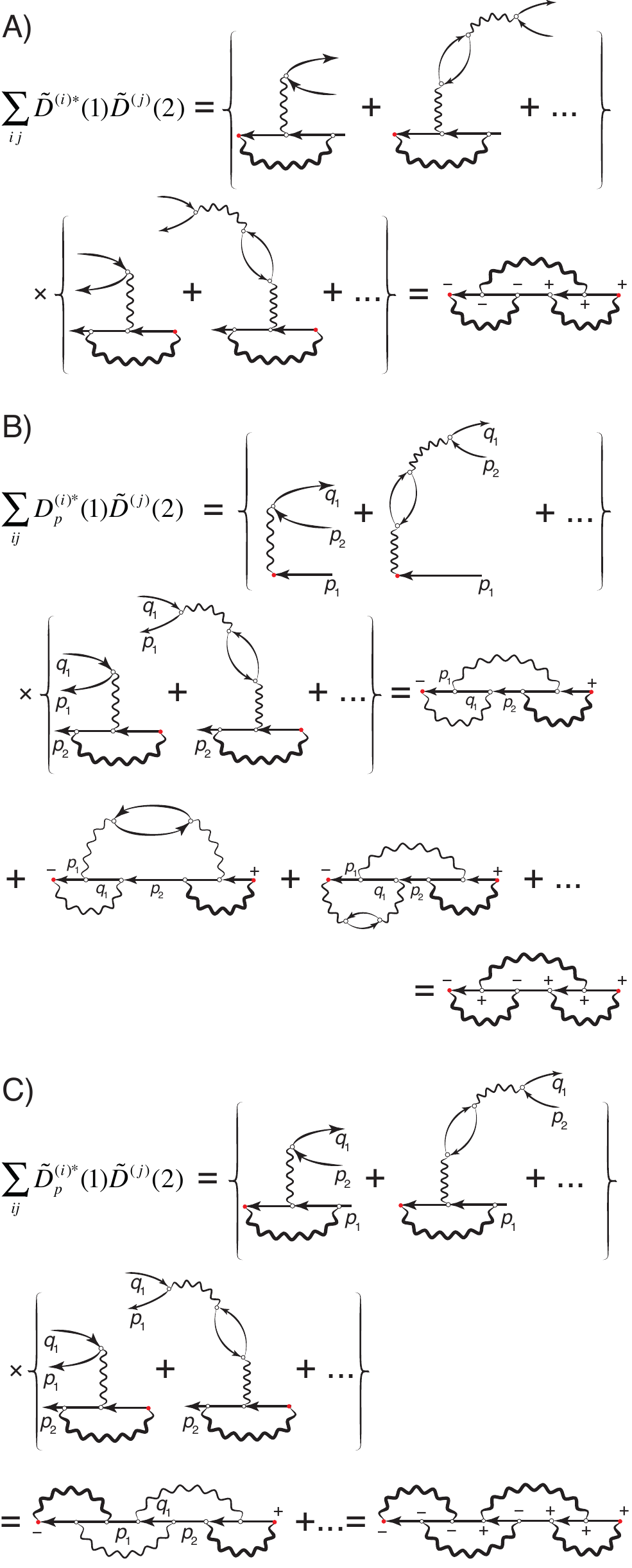}
\caption{(Color online) There are four extra diagrams that correct the self-energy with
  three $+/-$ $g$-lines ($\Si_{3}$). They have (A) $\tilde D(1) \tilde D(2)$, (B) $D_P(1)
  \tilde D(2)$, $\tilde D_P(1) D(2)$ (similar to diagram (B) and thus not shown), and
  (C) $\tilde D_P(1) \tilde D(2)$ structures. Subscript $P$ denotes a diagram
  with permuted $p$-indices.
\label{extradiags}}
\end{figure}
%---------------------------
This self-energy contains the original four partitions of $\Si_{3}$ plus four more
partitions arising from the permutation of the two $\pu$ dangling lines. The latter are
explicitly worked out in Fig.~\ref{extradiags}.
Collecting all results together we conclude that the leading order self-energy diagrams
with screened interactions and vertex corrections yielding a PSD spectral function are
those in Fig.~\ref{fig:fig10}. Here we recall that a diagram with no $+/-$ on the
internal vertices is the full diagram (hence the sum over all 
possible partitions). Noteworthy the minimal completion of the square 
requires a {\em fourth} order diagram in $W$. 
%----------------------------
\begin{figure}[t]
\centering
\includegraphics[width=0.9\columnwidth]{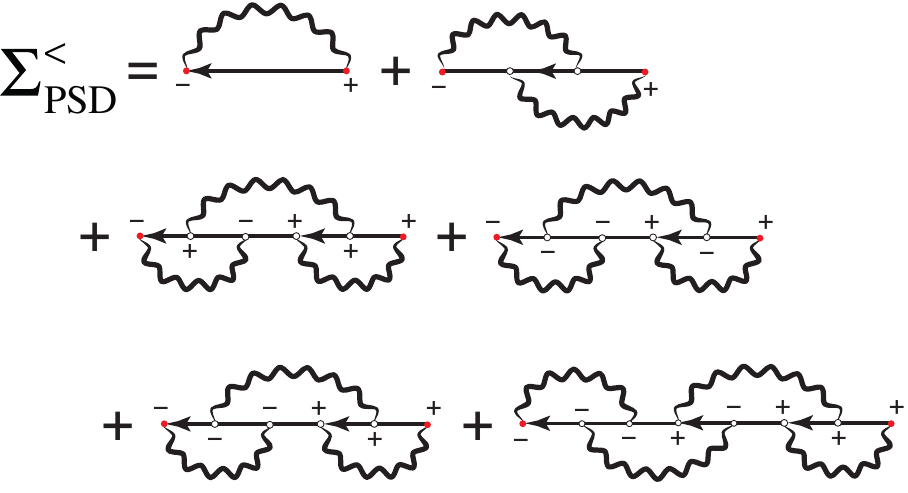}
\caption{(Color online) Leading order beyond-GW self-energy with the PSD property. Thick
  wavy lines denote the screened Coulomb interaction in the random phase approximation.}
\label{fig:fig10}
\end{figure}

%**************************************************************************************
\section{Vertex corrections in the homogeneous electron gas \label{sec:numerics}}
%**************************************************************************************

\begin{figure}[t!]
\centering
\includegraphics[width=0.9\columnwidth]{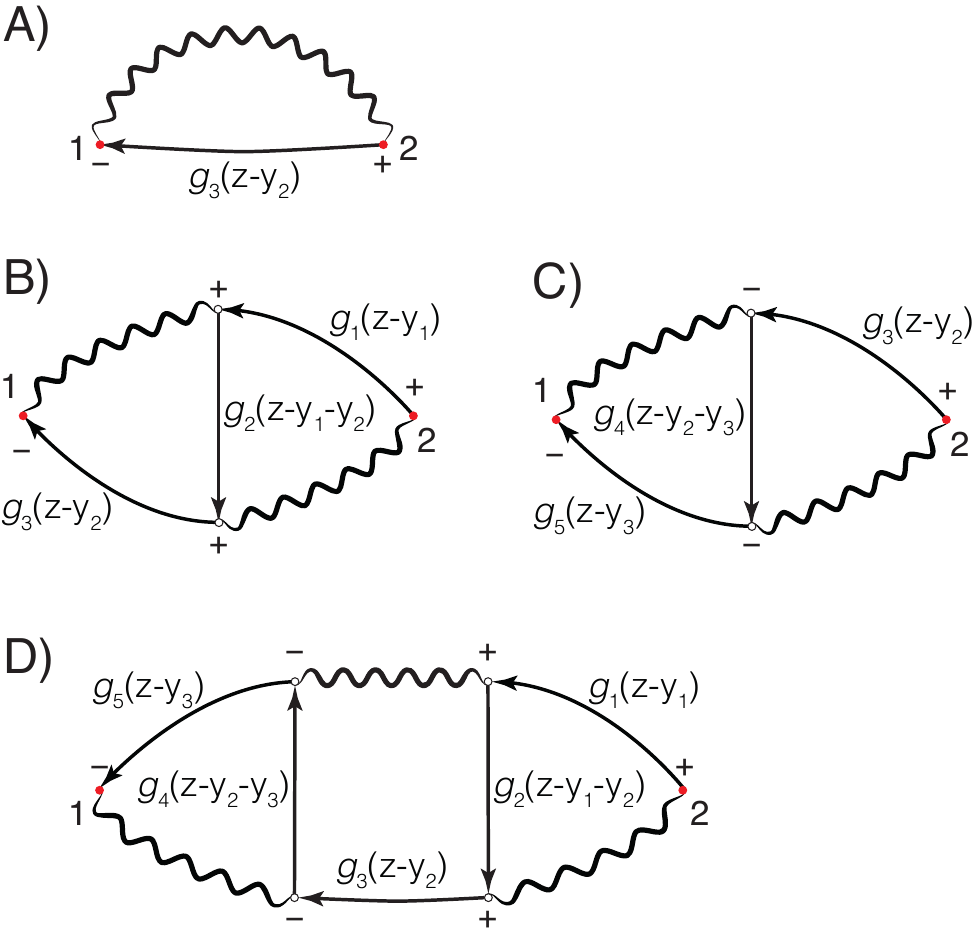}
\caption{The sum of the four self-energy partitions A)+B)+C)+D) yields a positive spectral
  function. Lines with arrows denote the electron propagator $G_{0}(k,\omega)$, whereas
  wavy-lines stand for the screened Coulomb interaction $W_{0}(k,\omega)$. Vertices are
  labeled with $+$ ($-$) if the time lies on the backward (forward) 
  branch  of
  the Keldysh contour. Diagrams are translated into analytic expressions according to the
  definitions \eqref{eq:g0} and \eqref{eq:W0} and notations below
  Eqs.~\eqref{eq:s123}. For instance $g_5(z-y_3)$, connecting vertices with 
  ``$-$'' labels as in C)
  and D), corresponds to the time-ordered Green's function (see also
  Eq.~(\ref{eq:gmm})):
  $\frac{\mathcal B(k_5)}{z-\Omega(y_3)-\epsilon_{k_5}-i\eta}+\frac{\mathcal
    A(k_5)}{z-\Omega(y_3)-\epsilon_{k_5}+i\eta}$.}
\label{fig:s123}
\end{figure}

The numerical implementation of the full self-energy of Fig.~\ref{fig:fig10} is rather
demanding. We observe, however, that the exclusion of the last partition of
Fig.~\ref{sigmavert} leads to a much simpler PSD self-energy since no permuted $D$
diagrams appear. It is straightforward to show that in this case the rules of
Eqs.~(\ref{rule1}-\ref{rule3}) lead to the following PSD self-energy
%===========================================================
%
%===========================================================
\bea
\Si^{<}_{c}(1,2)&=& i\sum_{j_{1}j_{2}=1}^{\infty}\sum_{\pu\qu}
\left( 
D^{(j_{2})}_{p_{1}p_{2}q_{1}}+\tilde{D}^{(j_{2})}_{p_{1}p_{2}q_{1}}\right)
\nn\\
&\times&
\left(
D^{(j_{1})^{\ast}}_{p_{1}p_{2}q_{1}}+\tilde{D}^{(j_{1})^{\ast}}_{p_{1}p_{2}q_{1}}\right),
\label{simplevertex}
\eea
%===========================================================
%
%===========================================================
where the corresponding $D$ diagrams are defined in Fig.~\ref{fig:g0w0}(A) and
Fig.~~\ref{fig:example4}(B).  The diagrammatic representation of Eq.~\eqref{simplevertex} is
shown in Fig.~\ref{fig:s123}. This self-energy too goes beyond the GW approximation, but
the vertex correction is only partial.

%----------------------------
%----------------------------

The three-dimensional homogeneous electron gas (3d HEG) is one of the most studied correlated
many-body system.\cite{giuliani_quantum_2005} We still lack detailed knowledge of one directly observable quantity --
the spectral function $A(k,\omega)$ -- when departing from the 
on-shell energy, i.e., when
$\omega\not\approx \epsilon_k$. Discrepancies with experimental measurements contributed
to the debates on the position of
satellites,~\cite{holm_self-consistent_1997,aryasetiawan_multiple_1996} bandwidth of
simple metals,~\cite{yasuhara_why_1999,takada_inclusion_2001} cancellation of vertex
function and self-consistency
effects,~\cite{minnhagen_aspects_1975,holm_fully_1998,schindlmayr_systematic_1998} and the
spectral function shape.~\cite{guzzo_valence_2011,pavlyukh_time_2013}

Despite numerous efforts there are just a few results that go beyond the GW approximation
in the study of quasiparticle properties. Analytically, these are the results of Onsager
\emph{et al.}~\cite{onsager_integrals_1966} on the second-order exchange energy and of
Ziesche~\cite{ziesche_self-energy_2007} and of Glasser and
Lamb~\cite{glasser_analysis_2007} on the second-order exchange self-energy. These works
only contain analytic results for on-shell self-energy and only a contribution of the bare
Coulomb interaction is included.  The screened Coulomb interaction is possible to treat
numerically. The non self-consistent calculations were performed by
Hedin~\cite{hedin_new_1965} and
Lundqvist.~\cite{lundqvist_single-particle_1967,lundqvist_single-particle_1967-1,lundqvist_single-particle_1968}
It took three decades to implement the same approach partially or fully
self-consistently. This was achieved in works by von Barth and
Holm.~\cite{von_barth_self-consistent_1996,holm_fully_1998} The non-positivity of the
spectral function first observed by
Minnhagen~\cite{minnhagen_vertex_1974,minnhagen_aspects_1975} hindered systematic
diagrammatic explorations and stimulated development of synthetic approaches: analyzing
real time Kadanoff-Baym dynamics,~\cite{kwong_real-time_2000} neglecting the incoherent
part of the electron spectral function,~\cite{shirley_self-consistent_1996} employing the
Ward identities and a model form of the exchange-correlation
kernel,~\cite{takada_inclusion_2001,takada_dynamical_2002,bruneval_many-body_2005,maebashi_analysis_2011}
or the self-consistent cumulant expansion.~\cite{holm_self-consistent_1997}

Using the presented formalism it is now possible to pursue the 
diagrammatic route. In this section we
present the results of a non self-consistent calculation.  Thus, we evaluate the diagrams in Fig.~\ref{fig:s123} for 3d HEG
using the analytical frequency and numerical Monte-Carlo momentum integrations. The method
was developed in a prior publication,~\cite{pavlyukh_time_2013} however, especially the
analytical frequency integration part had to be substantially extended.  For HEG the bare
time-ordered Green's function can be written as a function of frequency $\omega$ and
momentum $k$ as
%===========================================================
%
%===========================================================
\begin{equation}
G_{0}^{\minus\minus}(k,\omega)=\frac{\mathcal B(k)}
{\omega-\epsilon_k-i\eta}+\frac{\mathcal A(k)}{\omega-\epsilon_k+i\eta}.
\label{eq:g0}
\end{equation}
%===========================================================
%
%===========================================================
For self-consistent calculations this equation can be extended to include multiple poles
(e.g., to describe quasiparticle satellites) for each momentum value. In this work we
perform a one-shot calculation and therefore set $\mathcal B(k)=n_k$ and $\mathcal
A(k)=1-n_k$ with $0\le n_k\le 1$ denoting the occupation number and $\epsilon_k$ the energy
of the state with momentum $k$. For the energy dispersion we used the 
prescription by Hedin\cite{stefanucci_nonequilibrium_2013,hedin_new_1965,hedin_effects_1970}
to put the pole of the dressed 
$G=G_{0}+G_{0}\Si_{c}[G_{0}]G$ with Fermi momentum $k=k_{F}$ 
correctly at the Fermi surface, 
i.e., $\epsilon_k=k^2/2 +\mu-\epsilon_{F}$ where $\mu$ is the chemical 
potential and $\epsilon_{F}=k_{F}^{2}/2$ is the Fermi energy. Analogous 
definitions have been used for the anti-time-ordered, lesser and 
greater Green's functions.

Generally, each interaction line in Fig.~\ref{fig:s123} can designate either the bare
Coulomb interaction or include scattering with generation of a plasmon or a particle-hole
pair. On the RPA level the plasmon oscillator strength $t(q)$ vanishes when its dispersion
curve $\Omega(q)$ enters the particle-hole continuum at the critical wave number
$q_c$. Because plasmonic excitations exist on the bounded momentum interval they are
especially convenient for numerics: momentum integrations need to be performed on the
finite interval. Physically, plasmons also dominate the particle-hole response in electron
liquids with typical metallic densities. Therefore, in present work we only treat the
plasmonic contributions. The screened Coulomb interaction is treated in the plasmon
pole approximation, i.e. $W(k,\omega) \approx W_{0}(k,\omega)$ (cf. Eq.~25.11 of
Ref.~\onlinecite{hedin_effects_1970}):
%===========================================================
%
%===========================================================
\be
W_{0}^{--}(k,\omega)=\frac{v(k)}2\!
\left[\frac{w(k)}{\omega-\Omega(k)+i\eta}-\frac{w(k)}{\omega+\Omega(k)-i\eta}\right],
\label{eq:W0}
\ee
%===========================================================
%
%===========================================================
where we denote $w(q)=t(q) \Omega^2(0)/\Omega(q)$ and $0\le t(q)\le 1$ is the plasmonic
spectral weight with $t(0)=1$ and $t(q_c)=0$. Analogous defininitions 
have been used for the other Keldysh components of $W_{0}$.
Our numerical approach also allows us to
include contributions from the particle-hole continuum to exhaust the $f$-sum rule for the
dielectric function $\varepsilon(q,\omega)$:
%===========================================================
%
%===========================================================
\[
\int_0^\infty\omega\,\mathrm{Im}\varepsilon^{-1}(q,\omega)\,d\omega=-\frac\pi2\Omega^2(0),
\] 
%===========================================================
%
%===========================================================
where $\Omega(0)=4\sqrt{\frac{\alpha r_s}{3\pi}} \epsilon_F$, with 
$\alpha=[4/(9\pi)]^{1/3}\approx0.521$ and $r_s=1/(\a k_{F})$ the 
Wigner-Seitz radius.  Monte-Carlo momentum
integration of these terms is, however, more involved as it requires an extra integration for each
interaction line.  

The frequency integrations can be done completely analytically (facilitated by the {\sc
  mathematica} computer algebra system) whereas for the remaining momentum integrations
one has to rely on numerics. The
frequency integration is implemented for a general case of time-ordered
$G_{0}(k,\omega)$ and $W_{0}(k,\omega)$. Since each correlator in the self-energy
expression comprises two terms the total number of terms grows geometrically with the
diagrams order. In order to optimize the calculation we implemented an approach where for
each $\omega$-integration the program closes the integration contour in such a way that
the least number of terms is generated. One might argue that the Hedin set of equations
gives a prescription on which half 
of the complex plane the
integration should be closed (see for example Eqs.~(A30) of
Ref.~\onlinecite{hedin_new_1965}). However, it is easy to verify that this is only
relevant for the first order term. Higher order terms decay faster than $1/\omega$ at
infinity ($\omega$ here denotes the frequency to be integrated over) and, therefore, the
choice of the half-plane is not important. Finally, we notice that it is sufficient to consider
only the lesser self-energy. The greater self-energy component required to describe
properties of the states above the Fermi level is obtained from symmetry consideration.

Our analytic approach covers the case of Fig.~\ref{fig:s123}(A) diagram where all
partitions are included. Other diagrams on this figure contain only a subset of 
the partitions of the MBPT diagrams from which they originate. 
The corresponding analytic expressions can be easily extracted from the general
result by analyzing the frequency dependence. The overall $\omega$-dependence can readily
be obtained on paper by integrating $\delta$-functions in lesser and greater
correlators. Despite the omission of several partitions the final result is bulky and requires
an additional simplification using $\mathcal A(k)+\mathcal B(k)=1$.

\begin{widetext}
The results of frequency integration are
%===========================================================
%
%===========================================================
\begin{subequations}
\label{eq:s123}
\begin{eqnarray}
\Sigma_{A}^{<}(z,\zeta)&=&2i\pi\int\!\frac{d^3y_2}{(2\pi)^3}B_3C_2\delta(\zeta-\epsilon_3+\Omega_2),\label{eq:SA}\\
\Sigma_{B}^{<}(z,\zeta)&=&2i\pi\iint\!\frac{d^3y_1}{(2\pi)^3}\frac{d^3y_2}{(2\pi)^3}B_3C_1C_2
\left[\frac{\mathcal{H}_2(\epsilon_3,\Omega_1)-\mathcal{H}_1(\epsilon_3-\Omega_2,\Omega_1)}
{\epsilon_2-\epsilon_1-\Omega_2}\right]\delta(\zeta-\epsilon_3+\Omega_2),\label{eq:SB}\\
\Sigma_{C}^{<}(z,\zeta)&=&2i\pi\iint\!\frac{d^3y_3}{(2\pi)^3}\frac{d^3y_2}{(2\pi)^3}B_3C_2C_3
\left[\frac{\mathcal{H}_4(\epsilon_3,\Omega_3)-\mathcal{H}_5(\epsilon_3-\Omega_2,\Omega_3)}
{\epsilon_4-\epsilon_5-\Omega_2}\right]\delta(\zeta-\epsilon_3+\Omega_2),\label{eq:SC}\\
\Sigma_{D}^{<}(z,\zeta)&=&2i\pi\iiint\!\frac{d^3y_1}{(2\pi)^3}\frac{d^3y_2}{(2\pi)^3}\frac{d^3y_3}{(2\pi)^3}
B_3C_1C_2C_3
\left[\frac{\mathcal{H}_2(\epsilon_3,\Omega_1)-\mathcal{H}_1(\epsilon_3-\Omega_2,\Omega_1)}
{\epsilon_2-\epsilon_1-\Omega_2}\right]\nonumber\\
&&\quad\quad\quad\times
\left[\frac{\mathcal{H}_4(\epsilon_3,\Omega_3)-\mathcal{H}_5(\epsilon_3-\Omega_2,\Omega_3)}
{\epsilon_4-\epsilon_5-\Omega_2}\right]\
\delta(\zeta-\epsilon_3+\Omega_2),\label{eq:SD}
\end{eqnarray}
\end{subequations}
%===========================================================
%
%===========================================================
\end{widetext}
where we define 
%===========================================================
%
%===========================================================
\[
\mathcal{H}_i(a,\Omega)=\frac{A_i}{a-\Omega-\epsilon_i}+\frac{B_i}{a+\Omega-\epsilon_i}.
\]
%===========================================================
%
%===========================================================
In these equations we adopt the following notations: $A_i\equiv\mathcal A(x_i)$,
$B_i\equiv \mathcal B(x_i)$, $C_i\equiv \frac12v(y_i)w(y_i)$. The 
quantities $\epsilon_i$ and $\Omega_i$
are energies labeled by the momenta, as shown in Fig.~\ref{fig:s123}, 
and have a meaning
of electron and plasmon dispersions, respectively. The momenta $y_i$ are associated with
plasmonic excitations.  Eq.~\eqref{eq:s123} possesses some symmetries: Eqs.~\eqref{eq:SB}
and \eqref{eq:SC} are identical upon simultaneous permutations of the $(1,5)(2,4)$
fermionic and $(1,3)$ bosonic indices. The two brackets in Eq.~\eqref{eq:SD} transform
analogously.  For the numerical momentum integration of Eqs.~\eqref{eq:s123} it is useful
to rescale the variables as follows: $k=k_F\,\tilde k$,
$\epsilon_k=\epsilon_F\,\tilde{k}^2$,
$v(q)\equiv\frac{4\pi}{q^2}=\frac{4\pi}{k_F^2}\,\frac{1}{\tilde q^2}$,
$\Sigma=\epsilon_F\tilde{\Sigma}$. This
leads to the following density-dependent prefactors:
%===========================================================
%
%===========================================================
\begin{equation}
c_i=\left[\left(\frac{k_F}{2\pi}\right)^3\frac{1}{\epsilon_F^2}
\left(\frac12\frac{4\pi}{k_F^2}\epsilon_F\right)\right]^i=\left(\frac{\alpha r_s}{2\pi^2}\right)^i,
\label{eq:scale}
\end{equation}
%===========================================================
%
%===========================================================
where $i=1,2,3$ denotes the diagram's order.

From the particle-hole symmetry we can obtain the greater self-energy ($\Sigma^>$).  For
this it is sufficient to replace $B_i$ with $A_i$ and vice versa, and to change the sign
of each $\Omega_i$. 
\begin{figure}[h]
\centering
\includegraphics[width=0.9\columnwidth]{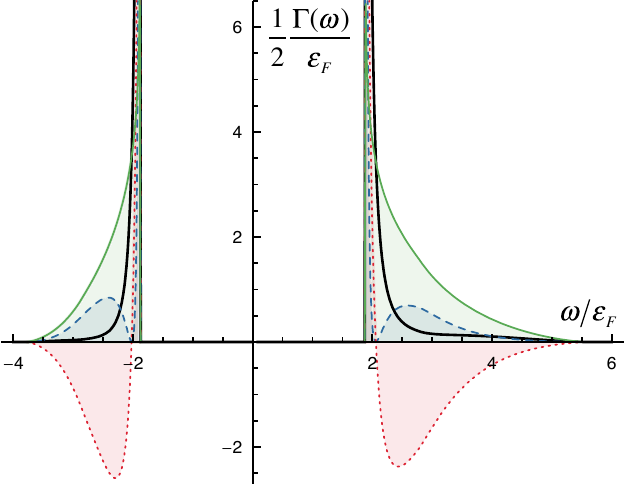}
\caption{(Color online) The rate operator
  $\frac12\Gamma(k,\omega)=-\Im\Sigma^{\rm R}_{c}(k,\omega)$ of the homogeneous electron gas
at the density of $r_s=4$ and $k=k_F$ (the energy $\w$ is measured 
with respect to $\mu$). Different line-styles denote contributions of
different orders: full, dotted and dashed lines stand for first, second and third order,
respectively. Thick solid line denotes the sum of all contributions.
 \label{fig:im_sgm_rs4_k1}}
\end{figure}
\begin{figure}[h]
\centering
\includegraphics[width=0.9\columnwidth]{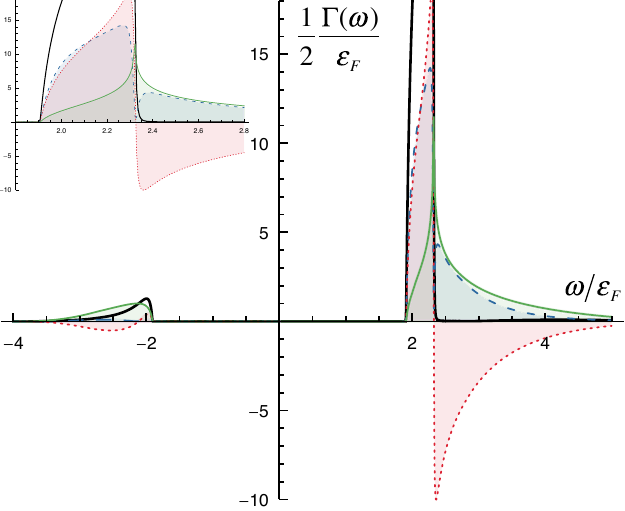}
\caption{(Color online) The rate operator 
$\frac12\Gamma(k,\omega)=-\Im\Sigma^{\rm R}_{c}(k,\omega)$
  as in Fig.~\ref{fig:im_sgm_rs4_k1} for a different momentum 
  $k=1.2k_F$ (the energy $\w$ is measured 
with respect to $\mu$). The inset magnifies
  the region of the logarithmic singularity. Notice the almost complete cancellation of
  $\Sigma^>_{c}(k,\omega)$ for $\omega>\epsilon_k+\Omega(0)$.
 \label{fig:im_sgm_rs4_k1p2}}
\end{figure}
\begin{figure}[t!]
\centering
\includegraphics[width=0.9\columnwidth]{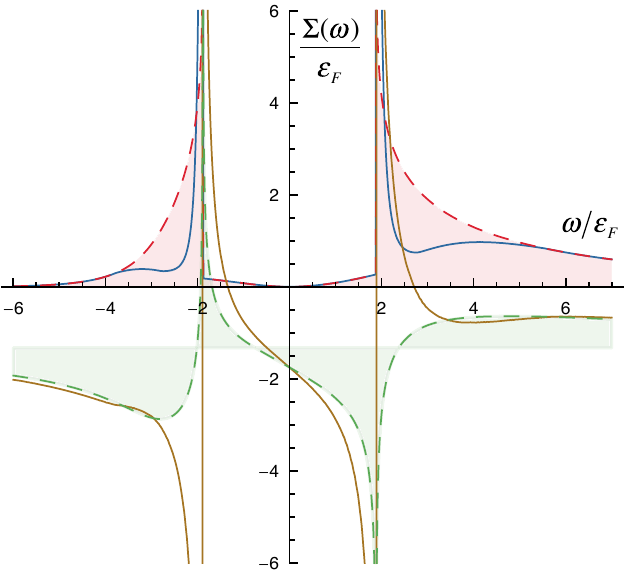}
\caption{(Color online) The real part of the retarded self-energy $\Re
  \Sigma^{\rm R}(k,\omega)$, see Eq. (\ref{sigx+c}), and the rate operator
  $\frac12\Gamma(k,\omega)=-\Im\Sigma^{\rm R}_{c}(k,\omega)$ of the homogeneous electron gas at the
  density of $r_s=4$ and $k=k_F$ (the energy $\w$ is measured 
with respect to $\mu$). Shaded dashed curves denote the first order
  calculation. Full lines denote contribution from diagrams shown in Fig.~\ref{fig:s123}
  plus the first order contribution from the particle-hole 
  excitations. The horizontal line bounding the shaded area in the 
  graph of $\Re
  \Sigma^{\rm R}$ crosses the 
  y-axis at the value of the exchange self-energy $\Sigma_x(k_{F})/\epsilon_F=-\frac{2\alpha}{\pi}
  r_s\approx -1.327$.
 \label{fig:sgm_rs4_k1}}
\end{figure}
\begin{widetext}
Upon a close inspection it is evident that the four terms in Eq.~\eqref{eq:s123} can be combined
together as follows:
%===========================================================
%
%===========================================================
\be
\Sigma_{2,\text{PSD}}^{<}=2i\pi\int\!\frac{d^3y_2}{(2\pi)^3}B_3C_2\left[
1+\int\!\frac{d^3y_1}{(2\pi)^3}C_1
\frac{\mathcal{H}_2(\epsilon_3,\Omega_1)-\mathcal{H}_1(\epsilon_3-\Omega_2,\Omega_1)}
{\epsilon_2-\epsilon_1-\Omega_2}\right]^2\delta(\zeta-\epsilon_3+\Omega_2).
\label{eq:fin}
\ee
%===========================================================
%
%===========================================================
\end{widetext}
The first order integrand (cf. Eq.~(\ref{eq:SA})) is multiplied by a full
square. Since we know that the GW self-energy fulfills the PSD property the same holds for the
sum of the four terms in Fig.~\ref{fig:s123}. This analytic conclusion is numerically
confirmed. 

In order to make Eqs.~\eqref{eq:s123} suitable for Monte-Carlo 
integration the following has
to be done: (i) integrate in spherical coordinates with zenith direction along the $\vec
z$ vector, (ii) one angular integration is trivially done ($2\pi$) because the system is
isotropic, (iii) map integration variables to the interval $[0,1]$. The speed and quality
of a pseudo-random number generator is very important for the present calculations. We use
the Mersenne twister 19937 generator as implemented in {\sc gsl} library combined with a
highly efficient jump ahead method~\footnote{implemented by K.-I. Ishikawa} due to
Haramoto \emph{et al.}~\cite{haramoto_efficient_2008} for parallelization. The method is
roughly 3 times faster than the standard {\sc fortran} implementation. We used roughly
$10^{12}$ Monte-Carlo realizations to get the full frequency dependence for each momentum
value. Real self-energy parts were computed using the Hilbert transform:
\be
\Sigma^{\rm R}(k,\omega)=\Si_{x}(k)+\int\frac{d\omega'}{2\pi}\frac{\Gamma(k,\omega')}{\omega-\omega'+i\eta},
\label{sigx+c}
\ee
where $\Si_{x}(k)$ is the frequency independent exchange 
self-energy and
$\Gamma(k,\omega)=i\big[\Sigma^>_{c}(k,\omega)-\Sigma^<_{c}(k,\omega)\big]
=i\big[\Sigma^{\rm R}_{c}(k,\omega)-\Sigma^{\rm A}_{c}(k,\omega)\big]$ is the rate operator.

Each term of Eq.~\eqref{eq:s123} was separately computed as shown in
Figs.~\ref{fig:im_sgm_rs4_k1},\ref{fig:im_sgm_rs4_k1p2} for two 
values of $k$.  The
first order self-energy is well understood. One of its marked
features is the existence of logarithmic singularities at
$\omega=\epsilon_k\pm\Omega(0)$.~\cite{hedin_effects_1970} These singularities have never
been observed in an experiment and are believed to be washed out by higher order
contributions. Although our calculations do not preclude this conclusion we found that our
selection of second-order terms makes the singularities even more pronounced.

The real part of the retarded self-energy 
of Eq.~(\ref{eq:fin}) is displayed in
Fig.~\ref{fig:sgm_rs4_k1}.
We observe that the real part of the first order and the complete
self-energy cross the y-axis at almost the same point which is equal 
to $\mu/\epsilon_{F}-1$.
This implies that the higher order corrections do not change the 
chemical potential appreciably.
As expected, the rate
operator is everywhere positive despite a large negative contribution of the second order
term. We notice an almost complete cancellations of different order terms beyond the
singularities, i.e. $\omega>\epsilon_k+\Omega(0)$ for particle ($k>k_F$) and
$\omega<\epsilon_k-\Omega(0)$ for hole ($k<k_F$) states. High accuracy of the Monte-Carlo
integration was required to get the cancellations properly. This is especially important
at metallic densities where different orders have comparable contributions. Due to the
density scaling (see Eq.~(\ref{eq:scale})) the first order self-energy becomes dominant at large
densities ($r_s\rightarrow 0$), while the third order is largest in the correlated low
density regime ($r_s\rightarrow\infty$).

The selection of diagrams of Fig.~\ref{fig:s123} and computed in this 
work in the plasmon pole
approximation describes scattering processes accompanied by the emission or absorption of
\emph{one} plasmon. Correspondingly, the scattering operator has pronounced (more narrow in
comparison with the first order result) peaks at $\omega=\epsilon_k\pm\Omega(0)$. But how
important are the remaining contributions?  The simplest first order term, absent in the
plasmon pole approximation, but included in the results of Fig.~\ref{fig:sgm_rs4_k1},
involves generation of a single particle-hole pair.  Due to less restrictions on the
available phase space for scattering it is important for $\omega\rightarrow\infty$ and also
determines the life-time of quasiparticles in the vicinity of the Fermi energy. It also
gives rise to secondary peaks in Fig.~\ref{fig:sgm_rs4_k1}.  These, however, are not
related to scattering mechanisms involving generation of two plasmons. Inclusion of such
processes is important for the interpretation of multiple satellites in the spectral
function. To lowest order they result from the partition of the second-order
self-energy diagram included in Fig.~\ref{fig:fig10}, but omitted in Fig.~\ref{fig:s123}.
Respective calculations are on the way and will be the subject of a forthcoming publication.

%**************************************************************************************
\section{Conclusions}
\label{c&osec}
%**************************************************************************************
Approximations of MBPT to the self-energy can lead to unphysical 
density of states, with a negative spectral weight in some frequency 
region. This undesired feature entails 
unphysical results on the system properties 
and makes self-consistent calculations impossible due to a 
progressive deterioration of the analytic properties of the Green's 
function. In 1985
Almbladh proposed a diagrammatic perturbation theory of the
photoemission current,~\cite{almbladh_theory_1985} and in a subsequent
paper~\cite{almbladh_photoemission_2006} he elaborated on the theory and gave a
prescription how to combine diagrams of different order to get a physically sensible
result: the positive-definite photoemission current. These ideas are precursors for our theory. Using
the Green's function of the Keldysh formalism we developed a method to construct
manifestly PSD spectral functions. The method becomes particularly lucid when expressed in
diagrammatic language as it amounts to apply a few simple drawing 
rules.

We derive a Lehmann-like representation of the exact self-energy
and show that it is given by the sum of squares of irreducible 
correlators.  We then elucidate the connection between the 
diagrammatic expansion of the irreducible 
correlators and MBPT. Any lesser/greater self-energy diagram can be partitioned into 
two halves with internal time-vertices on
opposite branches of the Keldysh contour. 
Thus, by simply drawing diagrams and assigning a sign to the internal 
vertices we are able  
to extend to a minimal set of diagrams any MBPT approximation and 
to generate PSD spectral functions.
Several important MBPT approximations, such as the GW or T-matrix
approximations, do not require any corrections. Our theory applies 
equally well to diagrammatic expansions with noninteracting and with 
self-consistent Green's functions  because PSD self-energies do preserve the correct analytic
structure. 

In standard MBPT approximations the straightforward inclusion of vertex corrections inevitably ruin 
the PSD property and, hence, our  additional diagrams must be included.
Remarkably, these diagrams are of higher order. For instance, the
inclusion of the \emph{full first-order vertex} leads to diagrams of the fourth order in
the screened interaction. Required computational power to numerically evaluate them is
immense. Fortunately, excluding some partitions allows us to construct an approximation
containing diagrams of maximally third order. They are feasible for numerics as our
calculations for the 3d HEG demonstrate.

Even though we only presented in detail the formalism  for 
the spectral function, the same ideas apply to the spectrum of 
the density response function. This extension, however, goes beyond 
the scope of the present work and will be presented elsewhere.

%**************************************************************************************
\section{Acknowledgments}
%**************************************************************************************
GS acknowledges funding by MIUR FIRB Grant
No. RBFR12SW0J. YP acknowledges support by DFG-SFB762. 
AMU would like to thank the Alfred Kordelin Foundation for support.
RvL would like to thank the Academy of Finland for support. 
%**************************************************************************************

%**************************************************************************************
%\bibliography{MyLibrary}

%**************************************************************************************
\end{document}